\documentclass[aps,showpacs,preprintnumbers,amsmath,amssymb,twocolumn,superscriptaddress,floatfix,nofootinbib,10pt,pre]{revtex4-1}

\usepackage[linesnumbered,lined,boxed,commentsnumbered,ruled,vlined]{algorithm2e}
\usepackage{algpseudocode}
\usepackage{amssymb, amsmath, amsfonts}
\usepackage{amsthm, mathrsfs, amsopn}
\usepackage{bm,bbm}
\usepackage{booktabs}
\usepackage{caption}
\usepackage{centernot}
\usepackage{colortbl}
\usepackage{comment}
\usepackage{dcolumn}
\usepackage{epstopdf}
\usepackage{float}
\usepackage{graphicx}
\usepackage[bottom]{footmisc}
\usepackage{mathrsfs}
\usepackage{mathtools}
\usepackage{multirow}
\usepackage{physics}
\usepackage{ragged2e}
\usepackage{subfigure}
\usepackage{tabularx}
\usepackage{tikz}
\usepackage{verbatim}
\usepackage{xkcdcolors}
\usepackage{hyperref}
\usepackage{cleveref}
\usepackage{amsthm}
\usepackage{mathrsfs}
\usepackage{mathrsfs}
\usepackage{blkarray}
\usepackage{nicematrix}
\DeclareCaptionJustification{justified}{\justifying}
\hypersetup{colorlinks=true, citecolor=orange, urlcolor=blue, linkcolor=magenta}

\begin{document}
\raggedbottom
\frenchspacing
\title[ ]{Global synchronization of topological signals with time-delayed interactions}
\author{Wilfried Segnou}
\affiliation{Department of Mathematics and Namur Institute for Complex Systems (naXys), University of Namur, Rue Graf\'e 2, B-5000 Namur, Belgium}
\affiliation{IMT School for Advanced Studies Lucca, Piazza San Francesco 19, 55100 Lucca, Italy}

\author{Thierry Njougouo}
\affiliation{IMT School for Advanced Studies Lucca, Piazza San Francesco 19, 55100 Lucca, Italy}

\author{Diego Garlaschelli}
\affiliation{IMT School for Advanced Studies Lucca, Piazza San Francesco 19, 55100 Lucca, Italy}
\affiliation{Lorentz Institute for Theoretical Physics, University of Leiden, Einsteinweg 55, 2333 CC, Leiden, The Netherlands}

\author{Timoteo Carletti}

\affiliation{Department of Mathematics and Namur Institute for Complex Systems (naXys), University of Namur, Rue Graf\'e 2, B-5000 Namur, Belgium}

\email{timoteo.carletti@unamur.be}
\email{wilfried.segnou@unamur.be}
\date{\today}

\begin{abstract}
Topological signals are dynamical variables supported on higher-order structures such as simplicial or cell complexes. In this work, we investigate the impact of time-delayed interactions on the emergence of global synchronization of topological oscillators, with application to the Stuart–Landau system. We first examine the case where signals are supported on simplexes, or cells, of a given  dimension and coupled through the Hodge-Laplace matrix. We derive the Master Stability Function and we prove that for suitable discrete delay values, the stability of the synchronous solution becomes analytically tractable and can be solved by using the Lambert W-function. This analysis yields explicit spectral conditions for Global Topological Synchronization to emerge and shows that admissible delays may either promote or suppress synchronization. We then study delayed interactions between signals supported on simplexes, or cells, of different dimensions coupled by the Dirac operator. Under analogous admissible  time delays the variational problem, still related to the Master Stability Function, can again be simplified, however in this case we prove that Global Topological Dirac Synchronization cannot emerge. Numerical simulations on simplicial and cell complexes corroborate and complement these findings.
\end{abstract}

\maketitle


\section{\label{sec:Introduction}Introduction}

Global synchronization is a ubiquitous collective phenomenon that arises in a broad range of real-world systems, including biological~\cite{ramirez2018modeling,walker1969acoustic,dumas2010inter}, technological~\cite{gambuzza2015memristor,buscarino2009experimental,megam2014implementing}, and mechanical systems~\cite{pantaleone2002synchronization,chopra2008synchronization}. To understand and characterize this behavior, early theoretical studies relied on graph-based models, where nodes represent dynamical units and edges encode pairwise interactions between them~\cite{ghosh2022synchronized,arenas2008synchronization}. Within this framework, spectral methods have provided a powerful way to elucidate the role of network structure on the emergence of collective behaviors such as global synchronization of identical oscillators~\cite{pecora1998master,mirollo1990synchronization}, cluster synchronization in partitioned graphs~\cite{gambuzza2019criterion,schaub2016graph}, and partial synchronization~\cite{terman1997partial,grave2007partial}.

However, pairwise interactions are often insufficient to faithfully describe the organization of many real systems. In recent years, higher-order structures such as simplicial complexes, cell complexes, and hypergraphs have emerged as more appropriate mathematical frameworks for modeling many-body interactions~\cite{giusti2016two,giusti2015clique,millan2025topology,bick2023higher,bianconi2021higher,battiston2026collective,roddenberry2022signal,hoppe2025don}. These structures allow us to capture interactions involving more than two units and may give rise to collective phenomena that are absent, or much less natural, in standard graph-based settings. In particular, growing attention has been devoted to topological signals, namely dynamical variables supported not only on nodes, but also on higher-dimensional simplexes or cells such as edges, triangles, and higher-dimensional faces. Such signals arise in several real contexts. For instance, in neuroscience, the activity of neuronal cell bodies may interact with synaptic activity and be directly affected by gliotransmission in the presence of brain tumors~\cite{araque2014gliotransmitters}; related ideas also appear in signal processing~\cite{calmon2023dirac}.

When topological signals themselves behave as oscillators, the underlying higher-order structure can strongly influence their collective dynamics. In particular, it has been shown that higher-order interactions may induce explosive synchronization and other nontrivial dynamical effects~\cite{millan2020explosive,calmon2022dirac,ghorbanchian2021higher}. More recently, attention has turned to Global Topological Synchronization (\textrm{GTS})~\cite{carletti2023global} and Global Topological Dirac Synchronization  (\textrm{GTDS})~\cite{carletti2025global}. These two phenomena describe global synchronization regimes for topological oscillators on simplicial or cell complexes. More precisely, \textrm{GTS} refers to the global synchronization of topological signals of the same dimension coupled through the Hodge-Laplacian, whereas \textrm{GTDS} describes global synchronization between signals supported on simplexes or cells of different dimensions and interacting through the discrete Dirac operator.

Despite these recent advances, the theoretical understanding of \textrm{GTS} and \textrm{GTDS} remains limited to settings in which interactions are instantaneous. In particular,
authors in~\cite{carletti2023global,carletti2025global} have shown that the emergence of both \textrm{GTS} and \textrm{GTDS} are constrained by nontrivial topological conditions on the underlying simplicial or cell complex, and may also require suitable weights~\cite{wang2024global}. However, by assuming that interactions occur instantaneously is a strong idealization of real dynamical systems. Indeed, in many applications, interactions are affected by finite transmission, processing, or propagation times, so that delays become an intrinsic part of the dynamics. This observation raises the natural question of whether \textrm{GTS} and \textrm{GTDS} can persist under time-delayed interactions, and how delay modifies their stability conditions.

Time delays are known to play a fundamental role in the dynamics of coupled oscillators. Over the last decades, substantial progress has been made in understanding delay-induced phenomena in oscillator networks, ranging from synchronization, chimera patterns and multistability to amplitude death and other collective dynamical effects~\cite{lee2009large,borner2020delay,danciu2021oscillator,yeung1999time,petit2015delay,bick2017robust,gjurchinovski2014amplitude,wetzel2022network}. These results show that, in many systems, delays are not merely perturbations but constitute essential mechanisms that shape collective behavior. Despite this large literature, the effect of delayed interactions on topological synchronization in higher-order structures remains unexplored.

In this work, we investigate the impact of a homogeneous delay on the emergence of \textrm{GTS} and \textrm{GTDS}. We first consider delayed interactions between topological signals of the same dimension. By using the Stuart-Landau (SL) system as a prototype model of coupled regular oscillators, we identify a class of admissible delays for which the variational dynamics becomes analytically tractable, allowing us to derive explicit stability conditions for \textrm{GTS}. By anticipating on the following, this class (see Eq.~\eqref{eq:condiionHodeLapTau}) select a discrete finite set of allowed delays, $\tau_p$, indexed by the integers $p$, depending on the coupling strength and the frequency of the SL limit cycle solution. Moreover, by means of a generalized order parameter, we numerically show that, in presence of delay, \textrm{GTS} occurs predominantly for delay values belonging to the above introduced class. We then turn to consider delayed interactions between SL oscillators supported on simplexes, or cells, of different dimensions coupled through the Dirac operator. In this setting, we show that the admissible class of delays given by Eq.~\eqref{eq:condiionHodeLapTau}, prevents the emergence of \textrm{GTDS}. 

Our analysis is grounded on the use of the Master Stability Function (MSF) applied to topological signals coupled via the Hodge-Laplace~\cite{carletti2023global} or Dirac~\cite{carletti2025global} coupling. By exploiting the Hodge or the Dirac decomposition we can study the stability of the synchronous solution by separating the influence of each mode, i.e., eigenvectors of the Hodge-Laplace matrix or the left and right singular vectors of the incidence matrix. Eventually to handle the presence of delay, that returns a transcendental characteristic equation, we resort to the Lambert W-function and its property to determine the critical values of the delays for the presence of synchronization or lack thereof. Our main conclusion is therefore that the effect of delay strongly depends on the coupling architecture: the admissible delays selected by Eq.~\eqref{eq:condiionHodeLapTau} can promote \textrm{GTS} under Hodge-Laplace coupling, whereas the same delays inhibit the emergence of \textrm{GTDS} under Dirac coupling.

The remainder of the paper is organized as follows. 
In Section~\ref{sec:char}, we recall the main definitions and mathematical tools necessary to study synchronization in topological signals supported on simplicial and cell complexes, namely, the incidence matrices, the associated Hodge Laplacians, and the discrete Dirac operator.  Section~\ref{sec2} is devoted to the study of delayed interactions between topological signals supported on simplexes or cells of the same dimension. We first introduce the general dynamical model where these signals are coupled through the Hodge-Laplacian and interact with a homogeneous time delay. In Subsection~\ref{sec:2A}, we specialize the general framework to Stuart-Landau oscillators and derive the variational equations governing perturbations around the globally synchronized topological state. The subsection~\ref{sec:2B} is then devoted to the numerical analysis of the delayed Hodge-Laplace case. The numerical results are presented and discussed in detail. Section~\ref{sec:4} addresses delayed interactions between topological signals supported on simplexes, or cells, of different dimensions and coupled through the discrete Dirac operator. We begin by introducing the general delayed Dirac-coupled dynamical model, which accounts for interactions between signals defined on adjacent dimensions of the underlying simplicial or cell complex. In Subsection~\ref{sec:4A}, this framework is applied once again to Stuart-Landau oscillators. We derive the corresponding variational dynamics and investigate analytically whether the admissible class of delays introduced in the Hodge-Laplace setting can support GTDS. Subsection~\ref{sec:4B} presents the associated numerical results and provides a detailed interpretation of the observed dynamical regimes. Finally, Section~\ref{sec:conclusion} summarizes the main analytical and numerical findings of the paper and discusses about perspectives.

\section{Definition of topological signals, simplicial and cell complexes}
\label{sec:char}

In this section, we briefly review the main definitions of topological signals defined on simplicial or cell complexes by introducing some basic operators used in the following analysis. Interested readers may consult the references~\cite{barbarossa2020topological,black2022hodge,calmon2023dirac,schaub2022signal,bianconi2021higher} for further details.

Given a non-negative integer $k$, a simplex of dimension $k$, for short a $k$-\emph{simplex}, is a collection of $(k+1)$ nodes
$\sigma^k = \{v_0, v_1, \dots, v_k\}$. A $(k-1)$-face of the $k$-simplex $\sigma^k$ is a simplex of dimension $(k-1)$ obtained by removing one node from $\sigma^k$. A \emph{simplicial complex} of dimension $K$, $\mathcal{X}$, is a finite collection of $k$-simplexes of dimension $k \le K$, being $K$ a non-negative integer, closed under inclusion of faces. In other words, if a simplex $\sigma^{(k)}$ belongs to $\mathcal{X}$, then all of its faces also belong to $\mathcal{X}$. A node is a $0$-simplex, an edge is a $1$-simplex, a triangle is a $2$-simplex, a tetrahedron is a $3$-simplex, and so on.  A given $k$-simplex $\sigma^k = \{v_0,\dots,v_k\}$ has exactly two possible orientations; once the latter has been fixed, we will denote the oriented $k$-simplex by
$[v_0,\dots,v_k]$.
Two orientations are said to be equivalent, or coherent, if one can be obtained from the other by applying an even number of transpositions, i.e., to exchange two elements. For example,  an edge formed by the nodes $v_0$ and $v_1$, possesses the two orientations: $[v_0,v_1] = -[v_1,v_0]$. For a triangle formed by the nodes $v_0,v_1,v_2$, we have
$[v_0,v_1,v_2] =[v_2,v_0,v_1] = [v_1,v_2,v_0]$, and $[v_0,v_1,v_2] = -[v_1,v_0,v_2]$.
Let us observe that orientation is different from directionality. An intuitive example would be to consider an electrical circuit in which each branch can be oriented arbitrarily in order to establish a convention for the movement of charges; a negative current then simply indicates that the actual flow is in the opposite direction to the chosen orientation. On the other hand, directionality would imply that charges can move only in a direction, such as in a diode. 

A $(k-1)$-face, $\sigma_j^{k-1}$, of a $k$-simplex $\sigma_i^k$ is called a boundary element, and we denote it by
$\sigma_j^{k-1} \subset \sigma_i^k$. Moreover we use the notation $\sigma_j^{k-1} \sim \sigma_i^k$ to indicate that the orientations of $\sigma_j^{k-1}$ and $\sigma_i^k$ are coherent, and $ \sigma_j^{k-1} \nsim \sigma_i^k$ to indicate that they are not coherent.
An oriented simplicial complex of dimension $K$ is completely characterized by the family of incidence matrices
$\mathbf{B}_k$, $k=1,\dots,K$, whose entries
encode whether a $(k-1)$-simplex belongs to the boundary of a $k-$simplex, with
the sign determined by their relative orientations, namely~\cite{bianconi2021higher,GradyPolimeni2010}
\begin{align}
\label{eq:incidenceMat}
 \mathbf{B}_k(i,j)=
\begin{cases}
0, & \text{if } \sigma_i^{k-1} \not\subset \sigma_j^k,\\[4pt]
1, & \text{if } \sigma_i^{k-1} \subset \sigma_j^k \text{ and }  \sigma_i^{k-1} \sim \sigma_j^k,\\[4pt]
-1, & \text{if } \sigma_i^{k-1} \subset \sigma_j^k \text{ and } \sigma_i^{k-1} \nsim \sigma_j^k\, .
\end{cases}
\end{align}
A simplicial complex can also be completely described by its Hodge-Laplace matrices~\cite{bianconi2021higher,torres2020simplicial,GradyPolimeni2010,lim2020hodge,horak2013spectra}
\begin{align}
\label{eq:HodgeLaplacian}
\mathbf{L}_0\text{ } &=  \mathbf{B}_1  \mathbf{B}_1^{\top}\,, \notag\\
\mathbf{L}_k\text{ } &= \mathbf{B}_k^{\top}  \mathbf{B}_k +  \mathbf{B}_{k+1}  \mathbf{B}_{k+1}^{\top}, \qquad k=1,\dots,K-1, \\
\text{ }\mathbf{L}_K &=  \mathbf{B}_K^{\top}  \mathbf{B}_K\notag\, .
\end{align}
For $k=1,\dots,K-1$, one can write $\mathbf{L}_k = \mathbf{L}_k^{\mathrm{down}} + \mathbf{L}_k^{\mathrm{up}}$, 
where $\mathbf{L}_k^{\mathrm{down}} =  \mathbf{B}_k^{\top} \mathbf{B}_k$ is the down Laplacian, encoding lower adjacency between $k$-simplexes, and $\mathbf{L}_k^{\mathrm{up}} =  \mathbf{B}_{k+1} \mathbf{B}_{k+1}^{\top}$
is the up Laplacian, encoding upper adjacency.

The Hodge-Laplace operators allow to encode adjacency relations among simplexes and can be used to couple topological signals defined on same dimension simplexes. To overcome this limitation and thus be able to couple topological signals of different dimensions, one can use the Dirac operator~\cite{bianconi2021topological,lloyd2016quantum,ameneyro2022quantum,post2009first}. The latter gathers all the incidence matrices of a simplicial or cell complex, and their transpose into a single matrix given by 
\begin{equation}
\label{eq:DiracOperator}
\mathbf{D}=
\begin{pmatrix}
0 & \mathbf{B}_1 & 0 & \cdots & 0\\
\mathbf{B}_1^\top & 0 & \mathbf{B}_2 & \ddots & \vdots\\
0 & \mathbf{B}_2^\top & 0 & \ddots & 0\\
\vdots & \ddots & \ddots & \ddots & \mathbf{B}_K\\
0 & \cdots & 0 & \mathbf{B}_K^\top & 0
\end{pmatrix}\, .
\end{equation}

\begin{figure}
    \centering
    \includegraphics[width=0.9\linewidth]{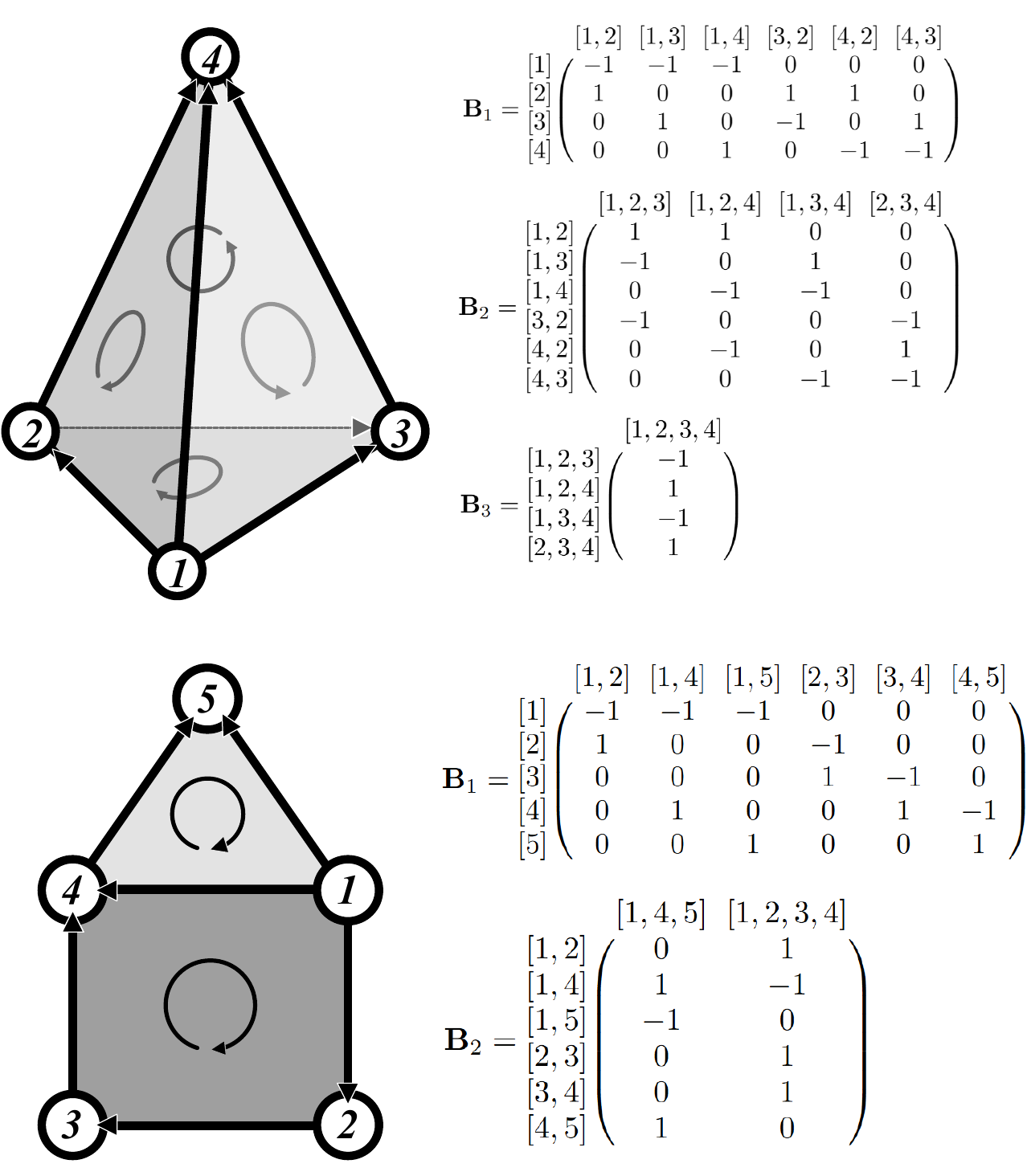}
    \caption{\textbf{$3$-simplicial complex and $2$-cell complex}. The top right panel shows a tetrahedron, i.e., a $3$-simplicial complex, while the top right panel displays the three associated incidence matrices. The bottom left panel, depicts a $2$-cell complex consisting of nodes, links, a triangle and a square, the bottom right panel reports the two associated incidence matrices. In both cases, edges orientations are denoted by the arrows drawn on the latter, while the faces by the curved arrows.}
\label{fig:Drawsimplicialcell}
\end{figure}

Cell complexes provide a natural generalization of simplicial complexes: instead of being built only from simplexes, they are built from more general cells, such as polygons, hypercubes, or other regular polytopes, while remaining closed under inclusion of faces. Most importantly, they carry the same algebraic-topological structure as simplicial complexes, since one can still define the incidence matrices, and the associated Hodge-Laplacians. In this sense, simplicial complexes appear as a particular class of cell complexes~\cite{roddenberry2022signal,hoppe2025don}. Figure~\ref{fig:Drawsimplicialcell} illustrates an example of a simplicial complex and a cell complex, along with their associated incidence matrices.
For a fixed dimension $k\leq K$, we consider topological signals supported on the $k$-simplexes of a simplicial complex, or more generally on the $k$-cells of a cell complex. If the structure contains $N_k$ simplexes or cells of dimension $k$, a $k$-topological signal assigns to each of them a vector in $\mathbb R^d$. We denote by $x(\sigma_i^k)\in\mathbb R^d$ the topological signal associated with the $i$-th $k$-simplex, or $k$-cell $\sigma_i^k$. Since simplexes and cells are oriented, the value of the signal depends on the chosen orientation. In particular, reversing the orientation of a simplex, or cell, changes the sign of the associated signal~\cite{carletti2023global}, namely
$x(\sigma_i^k)=-x(-\sigma_i^k)$. Throughout the paper, to lighten the notation, we will denote by $x_i^{(k)}$ the topological signal carried by the $i$-th $k$-simplex, or more generally by the $i$-th $k$-cell.
\begin{figure}
    \centering
    \includegraphics[width=0.9\linewidth]{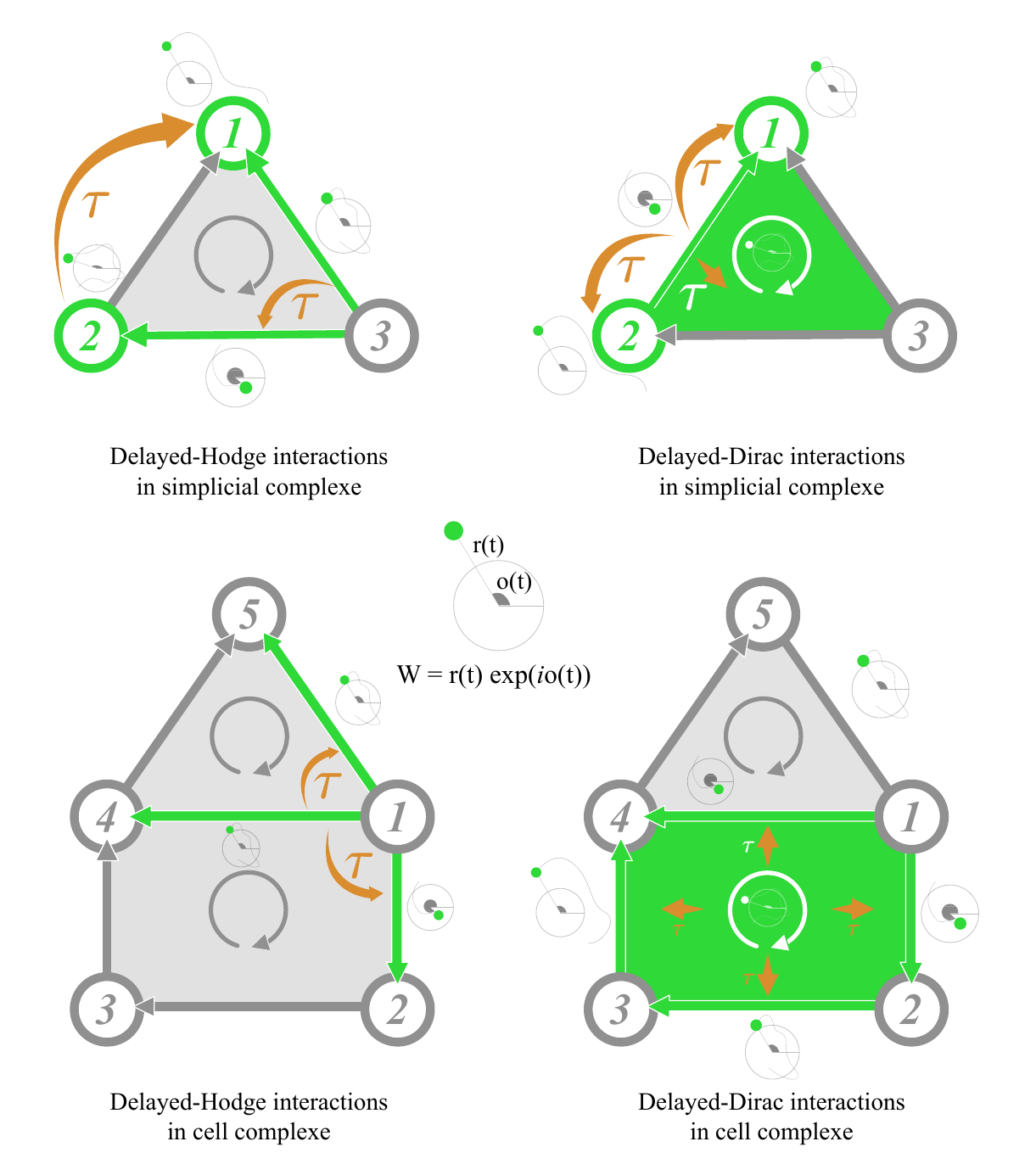}
    \caption{\textbf{Schematic representation of delayed Hodge interactions and delayed Dirac interactions}. Top panels illustrate delayed interactions mediated by the Hodge-Laplacian and the Dirac operator in a simplicial complex, whereas the bottom panels depict the corresponding interactions in a cell complex. Simplexes and cells are oriented, with straight arrows on links and curved arrows for $2$-simplexes or $2$-cells. Green highlights the simplexes, or cells, involved in each interaction, while orange arrows indicate how one simplex, or cell, influences the dynamics of another through an interaction with time delay $\tau$.
}
\label{fig:TopologicalsignalSimplicialandCell.}
\end{figure}

In the following sections, we will show how the Hodge-Laplacian and the Dirac operator can be used to describe delayed interactions between topological signals on simplicial and cell complexes. Figure~\ref{fig:TopologicalsignalSimplicialandCell.} provides an overview of such interactions, in which topological signals are defined by the complex amplitudes of limit-cycle oscillators. Top panels refer to the simplicial complex case, the toy example of a triangle. On the left, we illustrate the delayed interaction of the $0$-simplex $[2]$, with the $0$-simplex $[1]$, because of the presence of the $1$-simplex $[1,2]$; in the same panel, we also show the delayed interaction of the $1$-simplex $[1,3]$ with the $1$-simplex $[2,3]$, because they both belong to the $2$-simplex $[1,2,3]$, representing, i.e., the Hodge-Laplace delayed coupling. On the right, we show the $1$-simplex $[1,2]$ interacting with a time delay with the $0$-simplexes $[1]$ and $[2]$, and also with the $2$-simplex $[1,2,3]$, encoding, i.e., for the Dirac delayed coupling. The  bottom panels illustrate similar results for a cell complex. On the left, the $1$-simplex $[4,1]$ interacts with a delay with the $1$-simplex $[5,1]$ via the $2$-cell $[1,4,5]$ and the $1$-simplex $[1,2]$ via the $2$-cell $[1,2,3,4]$. On the right, the $2$-cell $[1,2,3,4]$ interacts with the links $[1,2]$, $[1,4]$, $[3,4]$, and $[3,2]$.

\section{Hodge-Laplace coupling with delay and Global Topological Synchronization}
\label{sec2}

In this section we consider the case of simplicial complexes, although the analysis can naturally be extended to cell complexes.
Let us consider a $K$-simplicial complex, fix $k\in\{0,\dots,K\}$ and consider the dynamics of a $k$-topological signal coupled to other $k$-topological signals through the $k$-th Hodge-Laplacian, $\mathbf{L}_k$, with a uniform time delay $\tau$, i.e., the same for all $k$-simplexes. The time evolution of the topological signals $x^{(k)}_i$, $i=1,\dots,N_k$, can thus be described by the following Delay Differential Equation (DDE)
\begin{equation}
\begin{aligned}
\frac{d}{dt}x^{(k)}_i(t)
=
f\bigl(x^{(k)}_i(t)\bigr)
-
\sum_{j=1}^{N_k} \mathbf{L}_k(i,j)\,h\bigl(x^{(k)}_j(t-\tau)\bigr)\, ,
\label{eq:delayed_hodge_general}
\end{aligned}
\end{equation}
where $f:\mathbb{R}^d\to\mathbb{R}^d$ is an odd nonlinear function that describes the intrinsic dynamics of the topological signal $x_i^{(k)}$ and $h:\mathbb{R}^d\to\mathbb{R}^d$ is an odd nonlinear coupling function.

\begin{figure}
    \centering
    \includegraphics[width=1\linewidth]{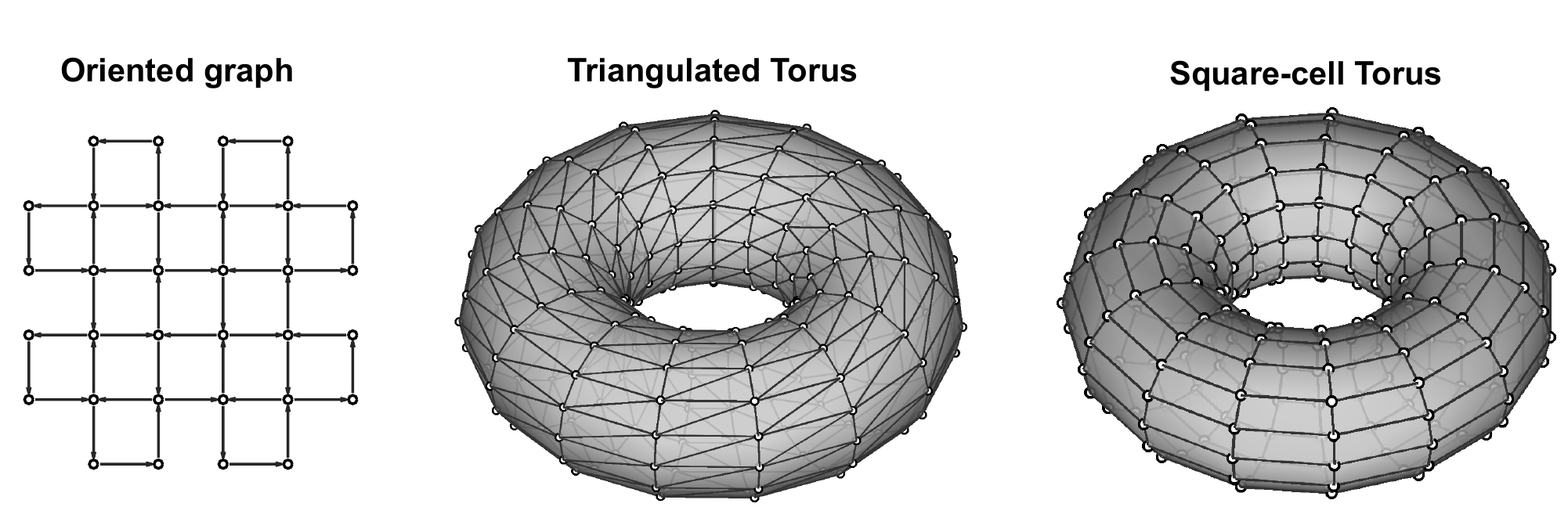}
    \caption{\textbf{Complexes used to illustrate the topological constraints for GTS.}. We present the structures used to perform the numerical results; on the left we have an oriented graph, i.e., a $1$-simplicial complex, in the middle the oriented triangulated torus, and on the right the oriented square-cell torus. We have deliberately omitted the orientations in the last two images for the sake of clarity.}
    \label{fig:illustrationUnderlyingnetwork}
\end{figure}

We are interested in global synchronization starting from a reference solution $s(t)\in\mathbb{R}^d$ of the uncoupled system
\begin{equation}
\label{eq:UncoupledDynamics}
\frac{ds}{dt}(t) = f(s(t))\, .
\end{equation}
A globally synchronized topological state of~\eqref{eq:delayed_hodge_general} is given by $x^{(k)}_i(t)={v}_i\,s(t)$ where $\mathbf{v}^{(k)}=(v_1,\dots,v_{N_k})^\top$ has entries in $\{-1,1\}$.  For such a state to exist, the sign vector $\mathbf{v}^{(k)}$ must belong to the kernel of the Hodge-Laplacian, namely $\mathbf{L}_k\mathbf{v}^{(k)}=0$. Because of~\eqref{eq:HodgeLaplacian} and the Hodge decomposition, this is equivalent to require $\mathbf{B}_k\mathbf{v}^{(k)} = 0$ and $\mathbf{B}_{k+1}^\top \mathbf{v}^{(k)} = 0$. The latter are well known as topological conditions for the emergence of \textrm{GTS} and have been extensively studied~\cite{carletti2023global}. The condition $\mathbf{B}_k\mathbf{v}^{(k)} = 0$ requires the orientations of the $(k-1)-$faces of the $k$-simplexes to be balanced, i.e., to have the same number of coherently and non coherently faces. The second condition, $\mathbf{B}_{k+1}^\top \mathbf{v}^{(k)} = 0$, can be satisfied only if $k$ is even, it thus imposes a geometrical constraint of the simplicial complexes that can sustain global synchronization. In the follow, we assume these topological conditions to hold true, and moreover we assume, up to reorient some simplex, that $\mathbf{v}^{(k)}=(1,1,\dots,1)^\top\in\mathbb{R}^{N_k}$; therefore, after a suitable reorientation of simplexes, all topological signals follow the same trajectory $s(t)$. Based on the above remark and for the sake of definitiveness, we decided to applied the proposed theory to the $1$-simplicial complex, the $2$-simplicial complex and the $2$-cell complex illustrated in Fig.~\ref{fig:illustrationUnderlyingnetwork}. For the oriented graph shown in the left panel, the $N_0$-dimensional uniform vector $\mathbf{v}^{(0)}$ always belongs to the kernel of the Hodge-Laplacian, $\mathbf{L}_0$, because every oriented edge has exactly one head and one tail node. Let us observe that, the chosen edge orientation is also balanced at every node, namely each node has the same number of incoming and outgoing edges. Consequently, the $N_1$-dimensional uniform vector also satisfies $\mathbf{L}_1\mathbf{v}^{(1)}=0$.  The triangulated torus, shown in the middle panel, admits the $N_0$ and $N_2$-dimensional uniform vectors in the kernels of the node and triangle Hodge-Laplacians, however the $N_1$-dimensional uniform vector does not belong to the kernel of the edge Hodge-Laplacian. More precisely, as in the graph case, the uniform node vector satisfies
$\mathbf{B}_1^\top\mathbf{v}^{(0)}=0$, and therefore $\mathbf{v}^{(0)}\in\ker(\mathbf{L}_0)$. Moreover, because the torus is a closed orientable surface, its triangles can be coherently oriented so that the contributions of the two triangles incident to each edge cancel, returning thus $\mathbf{B}_2\mathbf{v}^{(2)}=0$,
and then $\mathbf{v}^{(2)}\in\ker(\mathbf{L}_2)$.
The situation is different for edge signals. A uniform edge vector should to simultaneously satisfy $\mathbf{B}_1\mathbf{v}^{(1)}=0$ and $\mathbf{B}_2^\top\mathbf{v}^{(1)}=0$.
Although the edge orientations may be chosen so that the first condition is satisfied at every node, the second condition cannot hold true. Indeed, every triangle has three boundary edges and the corresponding component of $\mathbf{B}_2^\top\mathbf{v}^{(1)}$ is the sum of three incidence coefficients, each equal to $+1$ or $-1$. Since the sum of an odd number of terms in $\{-1,+1\}$ can never vanish, one necessarily has $\mathbf{B}_2^\top\mathbf{v}^{(1)}\neq 0$. In conclusion, $
\mathbf{v}^{(1)}\notin\ker(\mathbf{L}_1)$.
Notice that this does not imply that $\ker(\mathbf{L}_1)$ is trivial. The torus has nontrivial harmonic $1$-forms because of its topology, but these harmonic edge vectors are not spatially uniform. Finally, the square-cell torus, shown in the right panel, admits uniform vectors in the kernels of the Hodge-Laplacians in all three dimensions. 
The edge orientations are chosen so that the number of incoming and outgoing edges is balanced at every node, yielding $\mathbf{B}_1\mathbf{v}^{(1)}=0$.
 In contrast to a triangle, each square cell has four boundary edges. It is therefore possible to orient the edges so that two boundary incidences contribute with sign $+1$ and two with sign $-1$. Hence, $\mathbf{B}_2^\top\mathbf{v}^{(1)}=0$ and the uniform edge vector belongs to $\ker(\mathbf{L}_1)$. Moreover, the square cells can be coherently oriented because the torus is a closed orientable surface. The contributions of the two cells adjacent to each edge then cancel, giving $\mathbf{B}_2\mathbf{v}^{(2)}=0$, and consequently $\mathbf{v}^{(2)}\in\ker(\mathbf{L}_2)$.

To prove the existence of GTS we have to show the stability of the solution $x_i^{(k)}(t)=s(t)$ for all $i=1,\dots,N_k$. To achieve this goal we will follow the recipe of the Master Stability Function~\cite{fujisaka1983stability,pecora1998master} and we thus introduce heterogeneous perturbations about the synchronized trajectory $
\delta x^{(k)}_i(t)=x^{(k)}_i(t)-s(t)$. By linearizing Eq.~\eqref{eq:delayed_hodge_general} about the reference solution $s(t)$ we obtain
\begin{equation}
\begin{aligned}
\frac{d}{dt}\delta x^{(k)}_i(t)
&=
\mathbf{J}_f\bigl(s(t)\bigr)\,\delta x^{(k)}_i(t) +\\
&\quad -
\sum_{j=1}^{N_k}
\mathbf{L}_k(i,j)\,
\mathbf{J}_h\bigl(s(t-\tau)\bigr)\,\delta x^{(k)}_j(t-\tau),
\end{aligned}
\label{eq:linearized_hodge_system}
\end{equation}
where $\mathbf{J}_f$ and $\mathbf{J}_h$ denote the Jacobians of $f$ and $h$, respectively, evaluated on the reference solution. To make one step forward, we leverage on the symmetry of the Hodge-Laplacian, $\mathbf{L}_k$, and therefore the existence of an orthonormal eigenbasis
$\{\phi_\alpha^{(k)}\}_{\alpha=1}^{N_k}$ such that $\mathbf{L}_k \phi_\alpha^{(k)} = \Lambda_\alpha^{(k)} \phi_\alpha^{(k)}$, with $\Lambda_\alpha^{(k)}\geq 0$ denoting the $\alpha$-th eigenvalue.
By expanding the perturbations on this basis, $\delta x^{(k)}_i(t)
=\sum_{\alpha=1}^{N_k}
\delta \hat{x}^{(k)}_\alpha(t)\,\phi_\alpha^{(k)}(i)$, allows to decouple the linearized equation into independent modes
\begin{equation}
\frac{d}{dt}\delta \hat{x}^{(k)}_\alpha(t)
=
\mathbf{J}_f\bigl(s(t)\bigr)\delta \hat{x}^{(k)}_\alpha(t)
-
\Lambda_\alpha^{(k)}
\mathbf{J}_h\bigl(s(t-\tau)\bigr)
\delta \hat{x}^{(k)}_\alpha(t-\tau)\,,
\label{eq:modal_variational_equation}
\end{equation}
 which remains a nonautonomous equation and requires numerical methods to determine the stability of the synchronous solution. Note that, even in the special case where the resulting Jacobians are constant, we can seek solutions in the form $\delta \hat{x}^{(k)}_\alpha(t)=\,\hat{x}^{(k)}_\alpha(0)e^{\lambda t}$ and thus obtain the characteristic equation
\begin{equation}
    \det\!\left(
\lambda \mathbf{I}_d
-
\mathbf{J}_f
+
\Lambda_\alpha^{(k)} \mathbf{J}_h e^{-\lambda\tau}
\right)=0\,,
\end{equation}
which has an infinite number of solutions that are generally difficult to obtain analytically. The goal is to solve the previous $1$-parameter family of equations for $\lambda(\Lambda^{(\alpha)})$ and to determine conditions to ensure $\Re\left(\lambda(\Lambda^{(\alpha)})\right)<0$ for all modes $\alpha$ transverse to the synchronous manifold, i.e, $\alpha >1$. Let us observe the presence of the term, $e^{-\lambda\tau}$, due to the delay that returns a transcendental equation whose solution is, generally, not possible to be explicitly obtained by using standard algebraic methods. This highlights the main analytical difficulty introduced by time delays and motivate the search for special delay regimes in which the stability problem becomes analytically tractable.

In the rest of the Section we will consider in detail an application to Stuart-Landau oscillators, because of its special structure, we would be able to make some analytical progress.

\subsection{GTS of delayed coupled Stuart-Landau oscillators}
\label{sec:2A}

Let us now consider an explicit example by assuming the uncoupled dynamics to be governed by the Stuart-Landau system~\cite{kuramoto1984chemical,garcia2012complex,di2017benjamin}; the latter is the normal form of any nonlinear oscillator near a Hopf bifurcation, describing both the evolution of the amplitude and the phase, for this reason it has been largely used as benchmark system in problem of synchronization.

The delayed model Eq.~\eqref{eq:delayed_hodge_general} rewrites thus
\begin{equation}
\label{eq:SL_hodge_delayed}
\begin{aligned}
\frac{d}{dt} w_j^{(k)}(t)
&=
\sigma\, w_j^{(k)}(t)
-\beta\, w_j^{(k)}(t)\left\lvert w_j^{(k)}(t)\right\rvert^2 \\
&\quad
-\mu \sum_{\ell=1}^{N_k}
\mathbf{L}_k(j,\ell)\,
h\!\left(w_\ell^{(k)}(t-\tau)\right)\,,
\end{aligned}
\end{equation}
where $\sigma=\sigma_{\Re}+i\sigma_{\Im}$ and $
\beta=\beta_{\Re}+i\beta_{\Im}$ are complex model parameters, while 
$\mu=\mu_{\Re}+i\mu_{\Im}$ is the coupling strength whose functional form has been fixed as 
$h\left( z\right)=z^a \bar{z}^{a-1}= z \lvert z\rvert^{2a-2}$, with $a\in\mathbb{N}$. It has been shown that the linearized equation in the absence of time delay and considering the expression for $h(z)$, results in a constant Jacobian matrix, thereby facilitating the analytical analysis of global synchronization~\cite{segnou2026synchronization}. 
When $\sigma_{\Re}>0$ and $\beta_{\Re}>0$, the uncoupled Stuart-Landau system admits the stable limit-cycle solution
$s(t)=r_0 e^{i\omega t}$, where $r_0=\displaystyle\sqrt{\frac{\sigma_{\Re}}{\beta_{\Re}}}$ and $\omega=\sigma_{\Im}-\beta_{\Im}\displaystyle\frac{\sigma_{\Re}}{\beta_{\Re}}$.

Let us assume the simplicial complex to allow the vector $\mathbf{v}=(1,\dots,1)^\top\in\mathbb{R}^{N_k}$ in the kernel of the Hodge-Laplace matrix, hence $w_j^{(k)}=s(t)$ for $j=1,\dots,N_k$ is a solution of Eq.~\eqref{eq:SL_hodge_delayed}. To study its stability, we define a small perturbations of amplitudes and phases, $w_j^{(k)}(t)=s(t)\,(1+\rho_j^{(k)}(t))e^{i\theta_j^{(k)}(t)}$ with $|\rho_j^{(k)}|,\,|\theta_j^{(k)}|\ll 1$, adapted to the Stuart-Landau model. By inserting the latter relation in Eq.~\eqref{eq:SL_hodge_delayed}
 and by keeping only linear terms in $\rho_j^{(k)}$ and $\theta_j^{(k)}$, one obtains
\begin{widetext}
\begin{equation}
\left\{
\begin{aligned}
\frac{d}{dt}\, \rho_j^{(k)}(t)
&=
-2\sigma_{\Re}\,\rho_j^{(k)}(t)
-
|\mu|\,r_0^{m-1}
\sum_{\ell=1}^{N_k}
\mathbf{L}_k(j,\ell)
\Bigl[
m a_\tau\,\rho_\ell^{(k)}(t-\tau)
-
b_\tau\,\theta_\ell^{(k)}(t-\tau)
\Bigr],
\\[4pt]
\frac{d}{dt}\,\theta_j^{(k)}(t)
&=
-2\beta_{\Im}
\frac{\sigma_{\Re}}{\beta_{\Re}}\,
\rho_j^{(k)}(t)
-
|\mu|\,r_0^{m-1}
\sum_{\ell=1}^{N_k}
\mathbf{L}_k(j,\ell)
\Bigl[
m b_\tau\,\rho_\ell^{(k)}(t-\tau)
+
a_\tau\,\theta_\ell^{(k)}(t-\tau)
\Bigr],
\end{aligned}
\right.
\label{eq:linear_SL_amplitude_phase}
\end{equation}
\end{widetext}
where
$m=2a-1$, $a_\tau=\cos(\arg(\mu)-\omega\tau)$ and $b_\tau=\sin(\arg(\mu)-\omega\tau)$.

 By projecting $(\rho_j^{(k)},\theta_j^{(k)})$ onto the eigenvector $\phi_\alpha^{(k)}$ of $\mathbf{L}_k$, $\rho_j^{(k)}(t)=\sum_{\alpha=1}^{N_k}\hat\rho_\alpha^{(k)}(t)\phi_\alpha^{(k)}(j)$ and $\theta_j^{(k)}(t)=\sum_{\alpha=1}^{N_k}\hat\theta_\alpha^{(k)}(t)\phi_k^{(\alpha)}(j)$,
 we obtain the decoupled modal system
\begin{equation}
\frac{d}{dt}
\begin{pmatrix}
\hat\rho_\alpha^{(k)}(t)\\
\hat\theta_\alpha^{(k)}(t)
\end{pmatrix}
=
\mathbf{J}_f
\begin{pmatrix}
\hat\rho_\alpha^{(k)}(t)\\
\hat\theta_\alpha^{(k)}(t)
\end{pmatrix}
-
\Lambda_\alpha^{(k)} \mathbf{J}_h^\tau
\begin{pmatrix}
\hat\rho_\alpha^{(k)}(t-\tau)\\
\hat\theta_\alpha^{(k)}(t-\tau)
\end{pmatrix},
\label{eq:modal_SL_system}
\end{equation}
with
\begin{equation*}
\mathbf{J}_f=
\begin{pmatrix}
-2\sigma_{\Re} & 0\\[3pt]
-2\beta_{\Im}\dfrac{\sigma_{\Re}}{\beta_{\Re}} & 0
\end{pmatrix}
\text{ and } 
\mathbf{J}_h^\tau=
r_0^{m-1}|\mu|
\begin{pmatrix}
m a_\tau & -b_\tau\\[3pt]
m b_\tau & a_\tau
\end{pmatrix}\,.
\end{equation*}
Let us observe that, as already stated, the above matrices are constant ones even if they arise from the linearization about the time depending solution $s(t)$, this is due to the peculiar structure of the the Stuart-Landau model and the choice of the coupling function.

Then, by looking again for exponential growth at short time of the form
$\hat\rho_\alpha^{(k)}(t)=\hat\rho_\alpha^{(k)}(0)e^{\lambda t}$ and $\hat\theta_\alpha^{(k)}(t)=\hat\theta_\alpha^{(k)}(0) e^{\lambda t}$,
we obtain the characteristic equation $P_\alpha(\lambda)=0$, where $P_\alpha(\lambda)$ is given by
\begin{equation}
\begin{aligned}
P_\alpha(\lambda)
&= \lambda^2 +2\sigma_{\Re}\lambda
+ \Lambda_\alpha^{(k)} r_0^{m-1}|\mu|
\Biggl[
(m+1)a_\tau\,\lambda
+ 2\sigma_{\Re}a_\tau  \\
&
+ 2\beta_{\Im}
\frac{\sigma_{\Re}}{\beta_{\Re}}
\,b_\tau
\Biggr]
e^{-\lambda\tau} 
+ m\bigl(\Lambda_\alpha^{(k)} r_0^{m-1}|\mu|\bigr)^2
e^{-2\lambda\tau}\, .
\end{aligned}
\label{eq:quasipolynomial_general}
\end{equation}
The latter is a transcendental equation that cannot be generally solved in closed form; nevertheless, we can use the following assumption to simplify the above equation and improve  the analytical study
\begin{equation}
\label{eq:condba}
b_\tau=0 \text{ and } 
a_\tau=1\, .
\end{equation}
Observe that we discard the case $a_\tau=-1$ to avoid positive arguments of the Lambert $W$-function~\cite{CorlessEtAl1996,AslUlsoy2003,lehtonen2016lambert}, which would produce roots with positive real parts. In conclusion we have
\begin{equation}
\arg(\mu)-\omega\tau_p=2p\pi,
\qquad p\in\mathbb{Z}\,,
\label{eq:condiionHodeLapTau}
\end{equation}
let us notice that the condition $\tau_p>0$, imposes a constraint on $p$ once the parameters $\mu$ and $\omega$ have been defined.

Under this assumption Eq.~\eqref{eq:quasipolynomial_general} factorizes as
\begin{equation}
\begin{aligned}
\Bigl(
\lambda+2\sigma_{\Re}+m\Lambda_\alpha^{(k)}r_0^{m-1}|\mu|\,e^{-\lambda\tau_p}
\Bigr)
\Bigl(
\lambda+\Lambda_\alpha^{(k)}r_0^{m-1}|\mu|\,e^{-\lambda\tau_p}
\Bigr)\\
=0\, ,
\label{eq:factorized_characteristic}
\end{aligned}
\end{equation}
from which we can obtain two families of characteristic roots, each one indexed by $\kappa\in\mathbb{Z}$:
\begin{equation}
\lambda_{1,\kappa}^{(k,\alpha)}
=
-2\sigma_{\Re}
+
\frac{1}{\tau_p}
W_\kappa\!\left(
-m\Lambda_\alpha^{(k)}r_0^{m-1}|\mu|\,\tau_p\,e^{2\sigma_{\Re}\tau_p}
\right)\, ,
\label{eq:lambda_family_1}
\end{equation}
and
\begin{equation}
\lambda_{2,\kappa}^{(k,\alpha)}
=
\frac{1}{\tau_p}
W_\kappa\!\left(
-\Lambda_\alpha^{(k)}r_0^{m-1}|\mu|\,\tau_p
\right)\, ,
\label{eq:lambda_family_2}
\end{equation}
where $W_\kappa$ are the branches of the Lambert W-function. 
Therefore, \textrm{GTS} is linearly stable only if
\begin{equation*}
\Re(\lambda_{1,\kappa}^{(k,\alpha)})<0
\quad\text{and}\quad
\Re(\lambda_{2,\kappa}^{(k,\alpha)})<0\, ,
\end{equation*}
for every branch $\kappa\in\mathbb{Z}$ and for every non-zero eigenvalue $\Lambda_\alpha^{(k)}$. Since~\cite{shinozaki2006robust} $\max_{\kappa\in\mathbb Z}\Re( W_\kappa(z))=\Re(W_0(z))$, the asymptotic behavior of the modes of~\eqref{eq:modal_SL_system} is controlled by the principal branch $W_0$. Consequently, \textrm{GTS} occurs when
\begin{equation}
\label{eq:dispel}
\Re(\lambda(\Lambda^{(k)}_\alpha,\tau))=\max\{\Re(\lambda_{1,0}^{(k,\alpha)}),
\Re(\lambda_{2,0}^{(k,\alpha)})\}<0
\end{equation} 
for every non-zero eigenvalue $\Lambda_\alpha^{(k)}$. Let us observe that the left hand side of the latter equation defines the dispersion relation or Master Stability Function.

Let us first consider Eq.~\eqref{eq:lambda_family_2} for $\kappa=0$. Being the argument of the Lambert function negative, the condition $\Re(\lambda_{2,0}^{(k,\alpha)})<0$ is satisfied if and only if
\begin{equation}
\label{eq:condReLambda2}
    -\frac{\pi}{2}<-\Lambda_{\alpha}^{(k)}r_0^{m-1}|\mu|\,\tau_{p}<0\, .
\end{equation}
For fixed values of $\mu$ and $\omega$, we choose $\tau_p$ as a monotonically increasing discrete function of $p$, hence condition Eq.~\eqref{eq:condReLambda2} is satisfied for all $\Lambda_{\alpha}^{(k)}$ provided
\begin{equation}
\label{eq:condReLambda2b}
    0<\Lambda_{\max}^{(k)}r_0^{m-1}|\mu|\,\tau_{\mathrm{crit},2}<\frac{\pi}{2}\, ,
\end{equation}
where we introduced the critical delay, $\tau_{\mathrm{crit},2}>0$, such that the latter equation is satisfied for all $\tau_p < \tau_{\mathrm{crit},2}$.

Let now consider Eq.~\eqref{eq:lambda_family_1} with $\kappa=0$ and for a given $\tau_p$, define $\xi_c(\tau_p)<-\pi/2$ such that $\Re(W_0(\xi_c(\tau_p)))=2\sigma_{\Re}\tau_p$. Being the argument of the Lambert function again negative, the condition $\Re(\lambda_{1,0}^{(k,\alpha)})<0$ is satisfied if and only if
\begin{equation}
\label{eq:condReLambda1}
    \xi_c(\tau_p)<-m\Lambda_{\alpha}^{(k)}r_0^{m-1}|\mu|\,\tau_{p}e^{2\tau_p\sigma_{\Re}}<0\, .
\end{equation}
The latter relation can be written as
\begin{equation}
\label{eq:condReLambda1b}
    0<m\Lambda_{\alpha}^{(k)}r_0^{m-1}|\mu|\,\tau_{p}<-\xi_c(\tau_p)e^{-2\tau_p\sigma_{\Re}}\, ,
\end{equation}
and one can determine a second critical delay, $\tau_{\mathrm{crit},1}>0$ such that Eq.~\eqref{eq:condReLambda1b} is satisfied for all $0<\tau_p<\tau_{\mathrm{crit},1}$.

We have thus proved that both $\Re(\lambda_{1,0}^{(k,\alpha)})<0$ and $\Re(\lambda_{2,0}^{(k,\alpha)})<0$ are satisfied provided $0<\tau_p<\tau_{\mathrm{crit}}$, where $\tau_{\mathrm{crit}}=\min\{\tau_{\mathrm{crit},1},\tau_{\mathrm{crit},2}\}$.

We have thus established a critical-delay condition for a fixed simplicial or cell complex. More precisely, once the Hodge-Laplacian spectrum is fixed, both principal characteristic branches have negative real parts for values of $\tau_p$ smaller than $\tau_{\mathrm{crit}}$. 

We now adopt the complementary viewpoint. Instead of fixing the spectrum and varying the delay, we fix an admissible delay $\tau_p$ and determine the interval of Hodge-Laplacian eigenvalues for which the synchronized state is stable. This amounts to identifying a critical spectral value $\Lambda_c(\tau_p)$ such that the dispersion relation is negative for $ 0<\Lambda_{\alpha}^{(k)}<\Lambda_c(\tau_p)$, and becomes positive once $\Lambda_{\alpha}^{(k)}$ crosses this threshold. To establish the existence of such a stability boundary, let us first examine the asymptotic behavior of the two principal characteristic branches. Let us thus consider $\Lambda_\alpha^{(k)}|\mu|\simeq 0$. We can approximate Eqs.~\eqref{eq:lambda_family_1} and Eq.~\eqref{eq:lambda_family_2} by
\begin{equation*}
\lambda_{1,0}^{(k,\alpha)} \simeq -2\sigma_{\Re}
-m\Lambda_\alpha^{(k)} r_0^{m-1}|\mu|\,e^{2\sigma_{\Re}\tau_p}\, ,
\end{equation*}
and
\begin{equation*}
\lambda_{2,0}^{(k,\alpha)} \simeq
-\Lambda_\alpha^{(k)} r_0^{m-1}|\mu|\, ,    
\end{equation*}
both quantities are therefore negative. On the other hand, when
$\Lambda_\alpha^{(k)}|\mu| \to +\infty$, the corresponding branches satisfy
$
\Re\lambda_{1,0}^{(k,\alpha)} \to +\infty$ and $\Re
\lambda_{2,0}^{(k,\alpha)} \to +\infty$.
Hence, the system exhibits a stable region for small values of
$\Lambda_\alpha^{(k)}|\mu|$, and an unstable region for sufficiently large
values of these parameters.

An interesting question is then to determine the bifurcation point $\Re(\lambda(\Lambda^{(k)}_\alpha,\tau_p))=0$ at which the
dispersion relation changes sign and becomes positive, as well as how this
bifurcation point depends on the delay $\tau_p$. To this end, we solve $\Re(\lambda_{1,0}^{(k,\alpha)})=0$ and $\Re(\lambda_{2,0}^{(k,\alpha)})=0$ that give us from Eq.~\eqref{eq:lambda_family_1}
\begin{equation}
\label{eq:con1}
\Lambda_{\mathrm{bif}}^{(1)}
=
-\frac{2\sigma_{\Re}}
{m r_0^{m-1}|\mu|\cos(\Omega \tau_p)},
\end{equation}
for $\Omega\cot(\Omega\tau_p)=-2\sigma_{\Re}$  and $\Omega\tau_p\in\left(\frac{\pi}{2},\pi\right)$  since $\sigma_{\Re} >0$ and from Eq.~\eqref{eq:lambda_family_2}
\begin{equation}
\label{eq:con2}
\Lambda_{\mathrm{bif}}^{(2)}
=
\frac{\pi}{2r_0^{m-1}|\mu|\tau_p}.
\end{equation}
Thus, the bifurcation point corresponds to $\Lambda_c(\tau)=\min\{\Lambda_{\mathrm{bif}}^{(1)},\Lambda_{\mathrm{bif}}^{(2)}\}$. When $\tau_p$ becomes sufficiently large, $\Lambda_{\mathrm{bif}}^{(2)} \to 0$ and, consequently, the stability region tends to vanish. Thus, \textrm{GTS} emerges if the stability region is sufficiently large, induced by small values of $\tau_p$ such that all eigenvalues $\Lambda_\alpha^{(k)}$ lead to a negative dispersion relation. 

\subsection{Numerical results}
\label{sec:2B}

\begin{figure}[h!]
    \centering
    \includegraphics[width=0.8\linewidth]{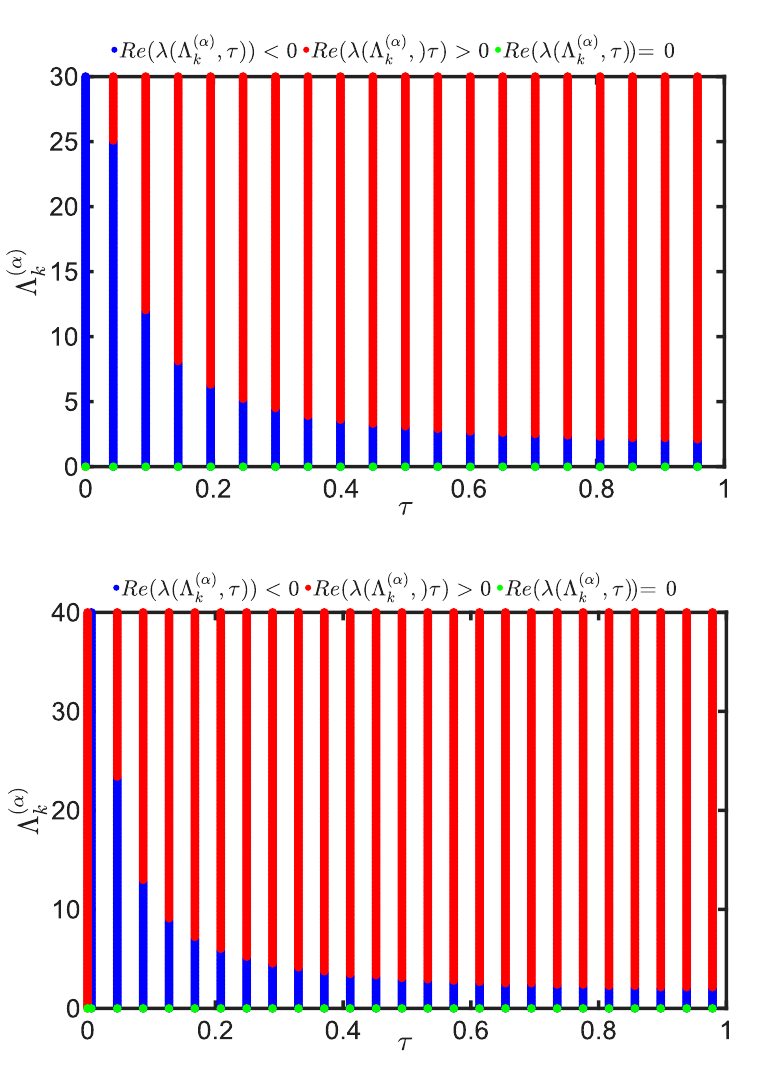}
    \caption{\textbf{Master Stability Function versus the admissible delay, $\tau_p$, and the eigenvalues of the Hodge-Laplace matrix, $\Lambda^{(k)}_\alpha$}. The top panel shows the sign of the MSF for eigenvalues $\Lambda^{(k)}_\alpha$ and discrete values of $\tau_p$ obtained with parameters $\mu = 0.1+0.5i$, $\sigma=1+2i$, $\beta = 1+50\pi i$ and $a=2$. Each vertical bar corresponds to the MSF as a function of $\Lambda^{(k)}_\alpha$ for a given $\tau_p$ obtained from Eq.~\eqref{eq:condiionHodeLapTau}. Blue indicates a negative MSF, while red indicates a positive MSF. The lower panel shows a similar result, this time with $\mu = 0.1-0.5i$.}
    \label{fig:msf_signs}
\end{figure}

For each admissible discrete value of the delay $\tau_p$ satisfying Eq.~\eqref{eq:condiionHodeLapTau}, there exists a stability window of eigenvalues close to  zero for which we thus have \textrm{GTS}. This behavior is summarized in Fig.~\ref{fig:msf_signs}. We choose a large value for the parameter, $\beta_{\Im}=50\pi$, so that several admissible values of $\tau_p$ lie in the interval $[0,1]$. In both panels of Fig.~\ref{fig:msf_signs}, every vertical bar represents the sign of the dispersion relation as a function of $\Lambda_\alpha^{(k)}$ for a fixed admissible value of $\tau_p$ (blue associated to negative values and red to positive ones). The leftmost vertical bar corresponds to the non-delayed case $\tau=0$. In the top panel, corresponding to the choice $\mu=0.1+0.5i$, returning thus $\beta_{\Im}\displaystyle\frac{\mu_{\Im}}{\beta_{\Re}}+\mu_{\Re}>0$, one can prove~\cite{segnou2026synchronization}  that the system achieves \textrm{GTS} as shown by the negativity of the MSF. As $\tau_p$ increases along the admissible discrete sequence, the stable region progressively shrinks and an unstable region, shown in red, appears for larger eigenvalues. The lower panel shows the complementary situation: for the corresponding parameter choice, $\mu=0.1-0.5i$, the non-delayed system has an unstable region near the origin, whereas a suitable, small enough, time delay can restore stability. These results suggest that time-delayed Hodge-Laplace interactions may either destroy or promote global topological synchronization, depending on the parameters and on the location of the Hodge-Laplacian spectrum.

\begin{figure}[!ht]
\centering \includegraphics[width=0.8\linewidth]{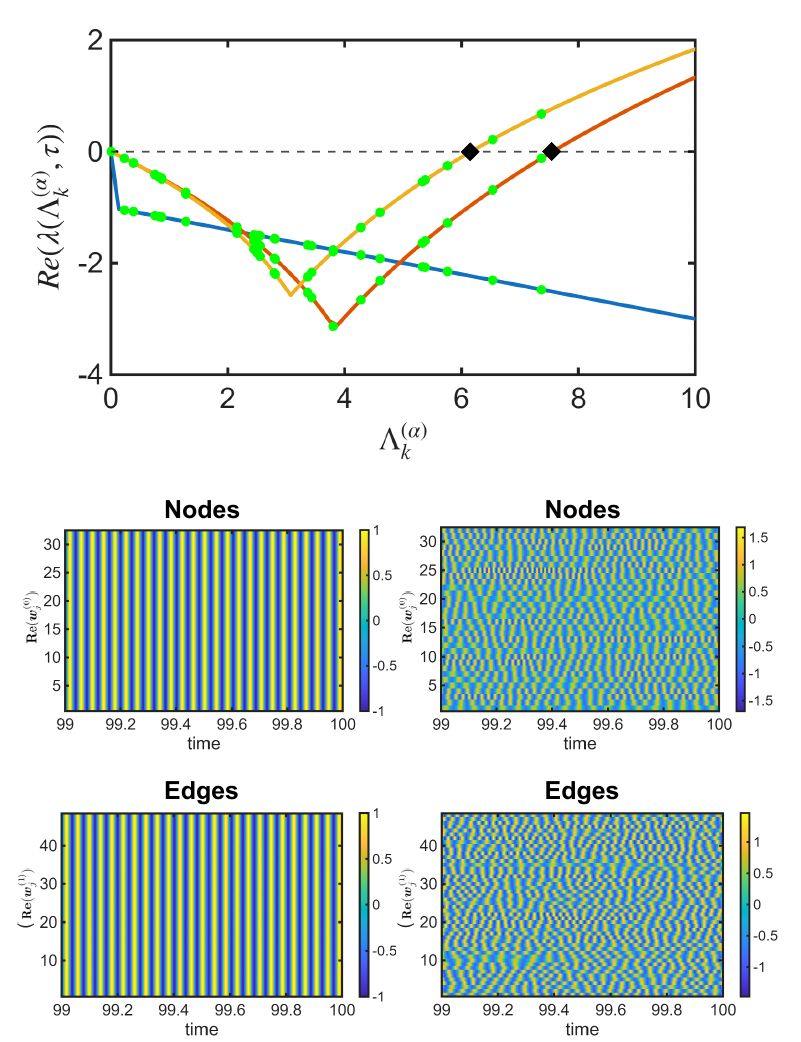} \caption{\textbf{Topological Desynchronization of Stuart-Landau oscillators induced by time delay}. Top panel displays the MSF for different values of $\tau$. The blue curve is obtained for $\tau=0$, whilst the red and orange curves are obtained respectively for $\tau_4 = (\arg{(\mu)}-2\cdot4\pi)/\omega$ and $\tau_5 = (\arg{(\mu)}-2\cdot5\pi)/\omega$, the model parameters are $\sigma = 1+2i$, $\beta = 1+50\pi i$, $\mu = 0.1+0.5i$ and $a=2$. The network (shown in the left panel of Fig.~\ref{fig:illustrationUnderlyingnetwork}) contains $N = 32$ nodes and $E = 48$ edges. The green points represent the eigenvalues of the Hodge-Laplacian matrix for the nodes and edges, both having the same nonzero eigenvalues. Black diamonds correspond to the bifurcation points calculated based on Eq.~\eqref{eq:con1} and Eq.~\eqref{eq:con2}. The middle and bottom panels show the time evolution of the real
part of the complex amplitude $w_j^{(k)}$ of the oscillators defined on nodes (middle panel) and on edges (bottom panel), i.e., for $k=0$ and $k=1$. The left column panels show the dynamics of the nodes and edges for $\tau_4 = (\arg{(\mu)}-2\cdot4\pi)/\omega$, whilst those on the right column do the same for $\tau_5 = (\arg{(\mu)}-2\cdot5\pi)/\omega$.} \label{fig:hodge_desynch_graph} 
\end{figure}

We illustrate the loss of synchronization induced by delay on a $1$-dimensional simplicial complex with $N=32$ nodes and $E=48$ edges (see left panel of Fig.~\ref{fig:illustrationUnderlyingnetwork} for a graphical representation). The node and edge Hodge-Laplacians,
$\mathbf{L}_0=\mathbf{B}_1\mathbf{B}_1^\top$ and $\mathbf{L}_1=\mathbf{B}_1^\top\mathbf{B}_1$, 
are isospectral and thus have the same nonzero eigenvalues. Top panel of Fig.~\ref{fig:hodge_desynch_graph} displays the MSF for three values of the delay, the blue curve corresponds to $\tau=0$, while the red and orange curves correspond to $\tau_4=\left(\operatorname{arg}(\mu)-2\cdot 4\pi\right)/\omega$ and $\tau_5=\left(\operatorname{arg}(\mu)-2\cdot 5\pi\right)/\omega$,
respectively. The green markers show the values of the dispersion relation evaluated at the eigenvalues of $\mathbf{L}_0$ and $\mathbf{L}_1$; the continuous curves are traced to help the visualization and are obtained by replacing the discrete eigenvalues with a continuous variable. For the parameters $\sigma=1+2i$, 
$\beta=1+50\pi i$, $\mu=0.1+0.5i$ and $a=2$, 
the MSF in the non-delayed case is negative, therefore, the \textrm{GTS} is linearly stable, provided the initial conditions are sufficiently close to the SL limit-cycle solution. For small admissible delays, such as $\tau_4$, all nonzero eigenvalues still lie in the stable part of the dispersion relation (see red curve), and synchronization is preserved (see panels in the left column). However, when the delay is increased to $\tau_5$, part of the spectrum enters the unstable region (see orange curve), leading to a loss of \textrm{GTS} (see panels in the right column). In the latter panels we indeed report the real part of the node dynamics (middle panels) and real part of the edge dynamics (bottom panels).

\begin{figure} 
\centering 
\includegraphics[width=0.8\linewidth]{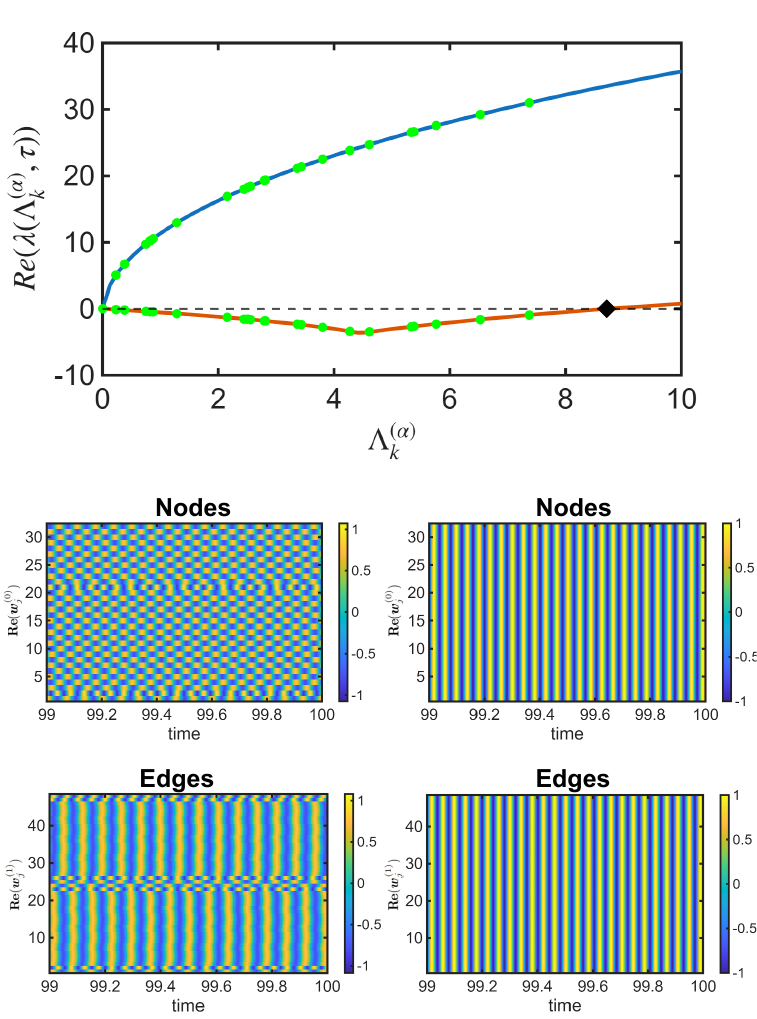} 
\caption{\textbf{Topological Synchronization of Stuart-Landau oscillators induced by time delay}. Top panel displays the MSF for two different values of $\tau$, the blue curve corresponds to $\tau=0$, whilst the red is obtained for $\tau_3 = (\arg{(\mu)}-2\cdot3\pi)/\omega$; the model parameters have been set to $\sigma = 1+2i$, $\beta = 1+50\pi i$ and $\mu = 0.1-0.5i$ and $a=2$. The black diamond  corresponds to the bifurcation point calculated from Eq.~\eqref{eq:con1} and Eq.~\eqref{eq:con2}. The left column panels show the time evolution of the real part of the complex amplitude of the oscillators defined on nodes (middle panel) and on edges (bottom panel) for $\tau = 0$, and one can clearly appreciate the absence of any regular behavior. The right column panels show the time evolution of the real part of the complex amplitude of the oscillators defined on nodes (middle panel) and on edges (bottom panel) for $\tau_3 = (\arg{(\mu)}-2\cdot3\pi)/\omega$, and now the synchronization is manifest.}
\label{fig:hodge_synch_graph} 
\end{figure}

We now consider a parameter regime in which time delay promotes synchronization. Specifically, we take
$\sigma=1+2i$, $\beta=1+50\pi i$, $\mu=0.1-0.5i$ and $a=2$. As shown in the top panel of Fig.~\ref{fig:hodge_synch_graph}, the non-delayed MSF is positive near the origin. Consequently, in the absence of delay, neither the node signals nor the edge signals synchronize. This is visible on the left column panels, where the real part of the complex amplitudes remain incoherent. In contrast, for the admissible delay
$\tau_3=\left(\operatorname{arg}(\mu)-2\cdot 3\pi\right)/\omega$,
the nonzero eigenvalues of both $\mathbf{L}_0$ and $\mathbf{L}_1$ fall inside the stable region of the MSF (see red curve in the top panel); this is confirmed by the results shown in the right column panels: both the node and edge signals synchronize when the time delay is introduced.

\begin{figure} \centering \includegraphics[width=0.8\linewidth]{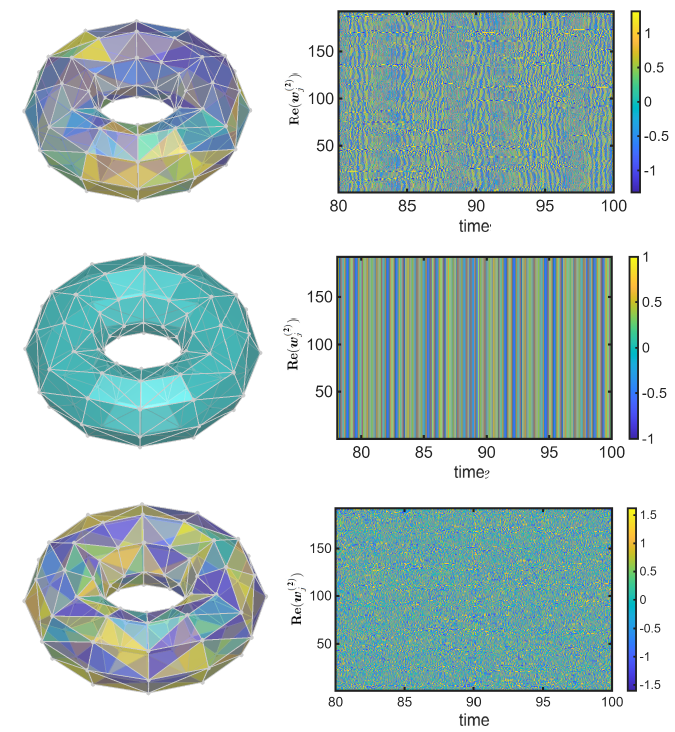} \caption{\textbf{Global Topological Synchronization of Stuart-Landau oscillators defined on $2$-simplicial complex}. We consider a $2D$-torus paved with $N_2 = 192$ $2$-simplexes each one hosting a Stuart-Landau oscillator and the coupled via the shared links. From top to bottom, the simulations are obtained with $\tau = 0$, $\tau_5=(\arg(\mu)-2\cdot5\pi)/\omega$ and $\tau_6=(\arg(\mu)-2\cdot6\pi)/\omega$. The remaining model parameters are $\sigma = 1 + 2i$, $\beta = 1 + 50\pi i$ and $\mu = 0.1-0.5i$.} 
\label{fig:hodge_synch_torus} 
\end{figure}

The results here reported, hold true beyond the example of the given $1$-simplicial complex. To support this claim we consider a $2$-dimensional simplicial complex, the $2D$-torus paved with triangles (see middle panel in Fig.~\ref{fig:illustrationUnderlyingnetwork}). Recall $2D$-torus paved with triangles cannot support GTS for $1$-topological signals. For this reason we consider the case of $2$-topological signals and the impact of delay. We fix a $2$-simplex with $N_2=192$ triangles, each one supporting a Stuart-Landau oscillator. The largest eigenvalue of the associated $2$-Hodge-Laplacian is $\Lambda_{192}^{(2)}=6$. For the choice of the model parameters $\sigma=1+2i$, $\beta=1+50\pi i$, $\mu=0.1-0.5i$ and $a=2$, the system without delay, i.e., $\tau=0$, does not synchronize, as shown in the top panels of Fig.~\ref{fig:hodge_synch_torus}. When the time delay is set to
$\tau_5=\left(\operatorname{arg}(\mu)-2\cdot 5\pi\right)/\omega$, all eigenvalues of the $2$-Hodge-Laplacian fall in the stable region of the dispersion relation, and the $2$-topological signals synchronize (see middle panels of Fig.~\ref{fig:hodge_synch_torus}). By further increasing the delay, for instance to $\tau_6=\left(\operatorname{arg}(\mu)-2\cdot 6\pi\right)/\omega$, part of the spectrum falls out of the stability region and synchronization is lost again (see bottom panels). These simulations confirm that time-delayed Hodge-Laplace coupling can either induce or suppress \textrm{GTS}, depending on the discrete time delay value selected by Eq.~\eqref{eq:condiionHodeLapTau} and the largest eigenvalue of $\mathbf{L}_k$. In other words, we can find a set of values for $\tau$ that allows the \textrm{GTS} to emerge for certain model parameters, whereas, for the same parameters, the absence of time delay does not allow the \textrm{GTS} to emerge.

To conclude this analysis we consider values of time delay not constrained by Eq.~\eqref{eq:condiionHodeLapTau}. To assess whether such delays promote or 
prevent \textrm{GTS}, we use the generalized order parameter
\begin{equation}
\label{eq:OP}
 R_{\tau}(t)=\frac{1}{r_0N_k}\left|\sum_{j=1}^{N_k}w_j^{(k)}(t)\right|   
\end{equation}
 
\begin{figure*} \centering \includegraphics[width=0.75\linewidth]{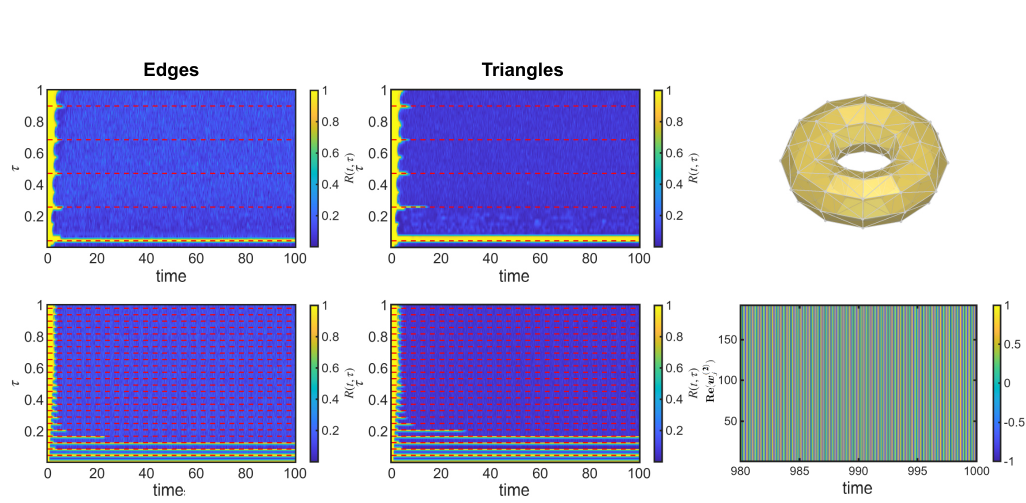} \caption{\textbf{Generalized order parameter}.
For the  network consisting of $32$ nodes and $48$ edges 
(the two leftmost panels), we compute the generalized order parameter 
$R(t,\tau)$ over time for values of $\tau$ in the interval $[0,1]$. 
The upper panel is obtained for $\beta = 1+10\pi i$, whereas the down
panel corresponds to $\beta = 1+50\pi i$. The red dashed lines indicate 
the evolution of the order parameter associated with the admissible 
discrete values $\tau_p$. The middle panels show the corresponding computation of the order 
parameter for the torus triangulation composed of $192$ triangles. The 
upper panel is obtained for $\beta = 1+10\pi i$, while the down panel 
corresponds to $\beta = 1+50\pi i$. The right panels show the dynamical 
integration of the simplexes defined on the triangles of the torus for 
$\tau=0.01$. The upeer panel displays the structure at the final 
time $t_f=1000$, whereas the bottom panel shows the time evolution 
of the real part of the complex amplitude $w_j$. The remaining parameters 
used in these simulations are 
$\sigma = 1+2i$, $\mu = 0.1-0.5i$, and $a=2$.} \label{fig:orderparameter} 
\end{figure*}

By varying $\tau$ in the interval $[0,1]$, we compute the order parameter as function of time, for both the network used in Figure~\ref{fig:hodge_desynch_graph} and the triangulated torus used in Figure~\ref{fig:hodge_synch_graph}. The results displayed in Figure~\ref{fig:orderparameter} show (left and middle panels) that synchronization can occur only for 
small admissible values of $\tau_p$ as one can appreciate by the value $R_{\tau_p}(t)= 1$ for all $t\geq 0$ (yellow). For the remaining values of the time delay \textrm{GTS} does not emerge, except in a few rare cases where $\beta_{\Im}$ 
is sufficiently large and $\tau$ remains very small. The right panels 
illustrate this situation for the dynamics defined on the triangles: the 
upper panel shows the time evolution of $\Re(w_j^{(k)})$ for $\tau=0.01\neq\tau_p$, whereas 
the bottom panel displays the torus structure at the final time $t_f=1000$.

\section{Dirac coupling with delay and Global Topological Dirac Synchronization}
\label{sec:4}

The Dirac operator allows to couple topological signals anchored to simplexes (or cells) of dimensions differing by one~\cite{carletti2025global,calmon2023dirac,zaid2026designing}, e.g., nodes with links, links with triangles and so on. Assume thus the topological signals associated with the simplexes (or cells) to have dimension $d$, and restrict attention to a $2$-dimensional simplicial, or cells, complexes. The same construction can be easily extended to higher-dimensional structures. 

Let $\mathbf{X}= ((x^{(0)})^\top,(x^{(1)})^\top,(x^{(2)})^\top)^\top$ be a topological spinor collecting all topological signals of the $2$-dimensional simplicial or cell complex $\mathcal{X}$ with $x^{(k)}=\big(x^{(k)}_1, x^{(k)}_2,\dots,x^{(k)}_{N_k}\big)$ for $k=0,1,2$.
The Dirac operator acts on the spinor $\mathbf{X}$ according to
\begin{equation}
\label{eq:Diracspinor}    
\mathbf{DX}=
\begin{pmatrix}
0 & \mathbf{B}_1 & 0\\
\mathbf{B}_1^\top & 0 & \mathbf{B}_2\\
0 & \mathbf{B}_2^\top & 0
\end{pmatrix}
\begin{pmatrix}
x^{(0)}\\
x^{(1)}\\
x^{(2)}
\end{pmatrix}
=
\begin{pmatrix}
\mathbf{B}_1x^{(1)}\\
\mathbf{B}_1^\top x^{(0)}+\mathbf{B}_2x^{(2)}\\
\mathbf{B}_2^\top x^{(1)}
\end{pmatrix}\, .
\end{equation}
Furthermore, we can split the Dirac operator as 
\begin{equation}
\label{eq: splitDirac}
\mathbf{D}=\mathbf{D}_1+\mathbf{D}_2\,,
\end{equation}
with $\mathbf{D}_1=
\begin{pmatrix}
0 & \mathbf{B}_1 & 0\\
\mathbf{B}_1^\top & 0 & 0\\
0 & 0 & 0
\end{pmatrix}$ and $\mathbf{D}_2=
\begin{pmatrix}
0 & 0 & 0\\
0 & 0 & \mathbf{B}_2\\
0 & \mathbf{B}_2^\top & 0
\end{pmatrix}$.

Then by using the Dirac decomposition~\cite{carletti2025global,calmon2023dirac,calmon2022dirac}, any topological spinor $\mathbf{X}$, can be uniquely written as the sum of three other topological spinors
\begin{equation}
\label{eq:spinordecomposition}
\mathbf{X}=\mathbf{X}_1+\mathbf{X}_2+\mathbf{X}_{\mathrm{harm}}\,,
\end{equation}
with
$\mathbf{X}_1\in\operatorname{im}(\mathbf{D}_1)$ is a topological spinor supported on nodes and edges,
$\mathbf{X}_2\in\operatorname{im}(\mathbf{D}_2)$ is a topological
spinor supported on edges and triangles and $\mathbf{X}_{\mathrm{harm}}\in\ker(\mathbf{D})$ is the harmonic component. Let us consider the vector field 
$\mathbf{F(X)}=\left(
f(x^{(0)})^\top,\\
f(x^{(1)})^\top,\\
f(x^{(2)})^\top
\right)^\top$ where each block acts componentwise, namely
$f(x^{(k)})=
\left(
f(x_1^{(k)}),\\
f(x_2^{(k)}),\\
\cdots\\
,f(x_{N_k}^{(k)}\right) 
$ and remember that the function $f$ must be odd to preserve the orientation invariance of the simplexes or cells. By introducing the operator
\begin{equation}
\label{eq:DiracCoupled}
\mathcal{D} =\gamma\begin{pmatrix}
0 & \mathbf{B}_1\otimes\mathbf{I}_d & 0\\
\mathbf{B}_1^\top\otimes\mathbf{I}_d & 0 & \mathbf{B}_2\otimes \mathbf{I}_d\\
0 & \mathbf{B}_2^\top\otimes \mathbf{I}_d & 0
\end{pmatrix}\, ,
\end{equation}
 where $\gamma$ is the coupling strength~\footnote{Let us observe that in the present case $\gamma$ is a real or complex number, the case where $\gamma\in \mathbb{R}^d\times\mathbb{R}^d$ has already been considered in~\cite{carletti2025global}.}, we deduce the general model describing the delayed dynamics of the topological spinor
\begin{equation}
\label{eq:Dynamicsspinor}
\frac{d\mathbf{X}}{dt}=\mathbf{F}(\mathbf{X})- \mathcal{D}\mathbf{H}_\tau (\mathbf{X}),
\end{equation}
where $\mathbf{H}_\tau(\mathbf{X})$ is the delayed coupling function defined as 
\begin{equation}
\label{eq:oddnonlinearCouplingfunction}
\mathbf{H_\tau(X)}=
\begin{pmatrix}
h(x^{(0)}(t-\tau))\\
h(x^{(1)}(t-\tau))\\
h(x^{(2)}(t-\tau))
\end{pmatrix}.
\end{equation}

Rewriting the model componentwise yields

\begin{align}
\label{eq: delayedmodel}
\begin{cases}
\displaystyle \frac{d}{dt}x_i^{(0)}
&=
f(x_i^{(0)})
-
\gamma
\displaystyle\sum_{\ell=1}^{N_1}
\mathbf{B}_1(i,\ell)\,
h\!\left(x_\ell^{(1)}(t-\tau)\right),\\
\displaystyle\frac{d}{dt}x_\ell^{(1)}
&=
f(x_\ell^{(1)})-
\gamma
\displaystyle\sum_{j=1}^{N_0}
\mathbf{B}_1^\top(\ell,j)\,
h\!\left(x_j^{(0)}(t-\tau)\right)+\\
&-
\gamma
\displaystyle\sum_{r=1}^{N_2}
\mathbf{B}_2(\ell,r)\,
h\!\left(x_r^{(2)}(t-\tau)\right), \\
\displaystyle\frac{d}{dt}x_r^{(2)}
&=
f(x_r^{(2)})
-
\gamma
\displaystyle\sum_{\ell=1}^{N_1}
\mathbf{B}_2^\top(r,\ell)\,
h\!\left(x_\ell^{(1)}(t-\tau)\right).
\end{cases}
\end{align}

Let us now consider a reference solution $s(t)\in \mathbb{R}^d$ of the isolated system $\displaystyle\frac{ds}{dt}=f(s)$. Global Topological Dirac Synchronization (\textrm{GTDS}) occurs when the topological spinor $\boldsymbol{\Phi}=\mathbf{v}\otimes s$ is a stable solution of Eq.~\eqref{eq:Dynamicsspinor},
where $\mathbf{v}=\left((\mathbf{v}^{(0)})^\top, (\mathbf{v}^{(1)})^\top, (\mathbf{v}^{(2)})^\top\right)^\top$ is a column vector with $\mathcal{N}=N_0+N_1+N_2$ entries, satisfying
$\mathbf{v}_i^{(0)}=1$ and $\mathbf{v}_i^{(k)}\in\{-1,1\}$ for  $0<k\le2$. A necessary condition for $\boldsymbol{\Phi}$ to be a solution of~\eqref{eq:Dynamicsspinor} is 
\begin{equation}
\label{eq:condB1B2}
\mathbf{B}_1\mathbf{v}^{(1)}=0,\quad
\mathbf{B}_2\mathbf{v}^{(2)}=0\text{ and }
\mathbf{B}_2^\top \mathbf{v}^{(1)}=0\, ,
\end{equation}
let us observe that $\mathbf{B}_1^\top \mathbf{v}^{(0)}=0$ is always satisfied.

In the following we will thus consider orientated simplicial or cell complexes for which Eq.~\eqref{eq:condB1B2} holds true with $\mathbf{v}=\mathbf{1}_\mathcal{N}=(\mathbf{1}_{N_0}^\top,\mathbf{1}^\top_{N_1},\mathbf{1}^\top_{N_2})^\top$. 

To investigate the stability of the synchronized solution $\boldsymbol{\Phi}=\mathbf{1}_\mathcal{N}\otimes s$ we consider again heterogeneous perturbations about the latter, $\delta \mathbf{X} =\mathbf{X}-\boldsymbol{\Phi}$, or in block form $\delta\mathbf{X}=\left((\delta x^{(0)})^\top,(\delta x^{(1)})^\top,(\delta x^{(2)})^\top
\right)^\top$, where $\delta x^{(k)}$ denotes the perturbation of the topological signal supported on the $k$-simplexes (or $k$-cells) about the synchronous solution $\mathbf{1}_{N_k}\otimes s$.

By linearizing Eq.~\eqref{eq:Dynamicsspinor} we obtain
\begin{equation}
\label{eq:GeneralizedLinearmodel}
\frac{d\,\delta\mathbf{X}}{dt}
=
\boldsymbol{\mathcal{J}}_f\,\delta\mathbf{X}
-
\mathcal{D}\,\boldsymbol{\mathcal{J}}_{h}^\tau\,\delta\mathbf{X}(t-\tau),
\end{equation}
where $\boldsymbol{\mathcal{J}}_f$ and $\boldsymbol{\mathcal{J}}_{h}^\tau$ are the Jacobian matrices of the local dynamics and the delayed coupling function, respectively. More precisely, these matrices are block diagonal and can be written as
\[
\boldsymbol{\mathcal{J}}_f=
\begin{pmatrix}
\mathbf{I}_{N_0}\otimes \mathbf{J}_f & 0 & 0\\
0 & \mathbf{I}_{N_1}\otimes \mathbf{J}_f & 0\\
0 & 0 & \mathbf{I}_{N_2}\otimes \mathbf{J}_f
\end{pmatrix},
\]
and
\[
\boldsymbol{\mathcal{J}}_{h}^\tau=
\begin{pmatrix}
\mathbf{I}_{N_0}\otimes \mathbf{J}_h^\tau & 0 & 0\\
0 & \mathbf{I}_{N_1}\otimes \mathbf{J}_h^\tau & 0\\
0 & 0 & \mathbf{I}_{N_2}\otimes \mathbf{J}_h^\tau
\end{pmatrix}
\]
where $\mathbf{J}_f$ is the Jacobian matrix of the function $f(x^{(k)}_i)$ evaluated at $s$, while $\mathbf{J}_h^\tau$ is the Jacobian matrix of the function ${h(x^{(k)}_i(t-\tau))}$. 
By invoking the Dirac decomposition, $\delta\mathbf{X}=\delta\mathbf{X}_1+\delta\mathbf{X}_2+\delta\mathbf{X}_{\mathrm{harm}}$, we can decouple the variational equations for the three components of the spinor, namely
\begin{equation}
\label{eq:DecoupledVariational}
\frac{d\,\delta\mathbf{X}_k}{dt}
=
\mathcal{J}_{f,k}\,\delta\mathbf{X}_k
-
\mathcal{D}_k\,\mathcal{J}_{h,k}^\tau\,\delta\mathbf{X}_k(t-\tau), \text{ with } k=1,2\, ,
\end{equation}
where we have defined
\begin{equation}
\label{eq:coupledD_1}
\begin{aligned}
\mathcal{D}_1 &=\gamma
\begin{pmatrix}
0 & \mathbf{B}_1\otimes \mathbf{I}_d & 0\\
\mathbf{B}_1^\top\otimes \mathbf{I}_d & 0 & 0\\
0 & 0 & 0
\end{pmatrix},
\mathcal{D}_2 =\gamma
\begin{pmatrix}
0 & 0 & 0\\
0 & 0 & \mathbf{B}_2\otimes \mathbf{I}_d\\
0 & \mathbf{B}_2^\top\otimes \mathbf{I}_d & 0
\end{pmatrix}\, ,
\end{aligned}
\end{equation}
and the matrices $\mathcal{J}_{f,k}$ and $\mathcal{J}_{h,k}^\tau$ are defined by
\begin{widetext}
\[
\mathcal{J}_{f,1}=
\begin{pmatrix}
\mathbf{I}_{N_0}\otimes \mathbf{J}_f & 0 & 0\\
0 & \mathbf{I}_{N_1}\otimes \mathbf{J}_f & 0\\
0 & 0 & 0
\end{pmatrix},
\mathcal{J}_{f,2}=
\begin{pmatrix}
0 & 0 & 0\\
0 & \mathbf{I}_{N_1}\otimes \mathbf{J}_f & 0\\
0 & 0 & \mathbf{I}_{N_2}\otimes \mathbf{J}_f
\end{pmatrix},
\]
and
\[
\mathcal{J}_{h,1}^\tau=
\begin{pmatrix}
\mathbf{I}_{N_0}\otimes \mathbf{J}_h^\tau & 0 & 0\\
0 & \mathbf{I}_{N_1}\otimes \mathbf{J}_h^\tau & 0\\
0 & 0 & 0
\end{pmatrix},
\mathcal{J}_{h,2}^\tau=
\begin{pmatrix}
0 & 0 & 0\\
0 & \mathbf{I}_{N_1}\otimes \mathbf{J}_h^\tau & 0\\
0 & 0 & \mathbf{I}_{N_2}\otimes \mathbf{J}_h^\tau
\end{pmatrix}.
\]
\end{widetext}

On the other hand, the harmonic component evolves according to the uncoupled linear equation
\setlength{\abovedisplayskip}{6pt}
\begin{equation}
\label{eq:linearequationharmoic}    
\frac{d}{dt}\delta \mathbf{X}_{\mathrm{harm}}
=
\boldsymbol{\mathcal{J}}_f\,\delta \mathbf{X}_{\mathrm{harm}}\, .
\end{equation}

The latter equation does not involve the coupling and thus neither the delay, hence GTDS depends on the stability of the solutions of Eq.~\eqref{eq:DecoupledVariational}. In the remaining part of this Section we consider again the Stuart-Landau model to present the theory above developed. 

\subsection{GTDS of delayed coupled Stuart-Landau oscillators}
\label{sec:4A}
For the sake of pedagogy, let us start by considering the case where topological signals are defined on edges and nodes of a graph, i.e., a $1$-simplicial complex; the system can thus be described by
\begin{equation}
\left\{
\begin{aligned}
\frac{d}{dt}u_j(t)
&=
\sigma\,u_j(t)-\beta\,u_j(t)\lvert u_j(t)\rvert^2
-\mu\sum_{\ell=1}^{N_1} \mathbf{B}_1(j,\ell)\,h(v_\ell(t-\tau)),
\\[4pt]
\frac{d}{dt}v_\ell(t)
&=
\sigma\,v_\ell(t)-\beta\,v_\ell(t)\lvert v_\ell(t)\rvert^2
-\mu\sum_{j=1}^{N_0} \mathbf{B}_1^\top(\ell,j)\,h(u_j(t-\tau)),
\end{aligned}
\right.
\label{eq:SL_dirac_delayedgraph}
\end{equation}
where $u_j$ and $v_\ell$ are the complex amplitudes defined on the nodes and edges respectively, $h(z) = z|z|^{2a-2}$ is the coupling function. We now consider synchronization solution $\boldsymbol{\Phi}=\left(\mathbf{1}_{N_0+N_1}\right)^\top\otimes s$ where $s(t)=r_0 e^{i\omega t}$, $r_0=\displaystyle\sqrt{\frac{\sigma_{\Re}}{\beta_{\Re}}}$ and $\omega=\sigma_{\Im}-\beta_{\Im}\displaystyle\frac{\sigma_{\Re}}{\beta_{\Re}}$, is the stable limit-cycle solution of the SL system. To study the stability of the latter we again linearize~\eqref{eq:SL_dirac_delayedgraph} about the synchronization solution by introducing heterogeneous perturbations in amplitude and phase of topological signals by writing $u_j(t)=\mathbf{s}(t)\bigl(1+\rho_j(t)\bigr)e^{i\theta_j(t)}$ and $v_\ell(t)=\mathbf{s}(t)\bigl(1+\eta_\ell(t)\bigr)e^{i\varphi_\ell(t)}$ where $\rho_j$, $\theta_j$, $\eta_\ell$ and $\varphi_\ell$ are small real-valued perturbations.
By substituting these expressions into the delayed equations and retaining only the linear terms yields
\begin{widetext}
{\small

    \begin{equation}
\left\{
\begin{aligned}
\frac{d}{dt}\rho_j
&=
-2\sigma_{\Re}\rho_j
-|\mu|r_0^{m-1}\sum_{\ell=1}^{N_1}\mathbf{B}_1(j,\ell)
\Bigl[
ma_\tau\,\eta^\tau_\ell-b_\tau\,\varphi^\tau_\ell
\Bigr],
\\[4pt]
\frac{d}{dt}\theta_j
&=
-2\beta_{\Im}\frac{\sigma_{\Re}}{\beta_{\Re}}\rho_j
-|\mu|r_0^{m-1}\sum_{\ell=1}^{N_1}\mathbf{B}_1(j,\ell)
\Bigl[
mb_\tau\,\eta^\tau_\ell+a_\tau\,\varphi^\tau_\ell
\Bigr],
\\[4pt]
\frac{d}{dt}\eta_\ell
&=
-2\sigma_{\Re}\eta_\ell
-|\mu|r_0^{m-1}\sum_{j=1}^{N_0} \mathbf{B}_1^\top(\ell,j)
\Bigl[
ma_\tau\,\rho^\tau_j-b_\tau\,\theta^\tau_j
\Bigr],
\\[4pt]
\frac{d}{dt}\varphi_\ell
&=
-2\beta_{\Im}\frac{\sigma_{\Re}}{\beta_{\Re}}\eta_\ell-|\mu|r_0^{m-1}\sum_{j=1}^{N_0} \mathbf{B}_1^\top(\ell,j)
\Bigl[
mb_\tau\,\rho^\tau_j+a_\tau\,\theta^\tau_j
\Bigr],
\end{aligned}
\right.
\label{eq:linear_nodes}
\end{equation}}
\end{widetext}
where we use the notation $x_j^\tau = x_j(t-\tau)$ and where $a_\tau=\cos(\arg(\mu)-\omega\tau)$ and  $b_\tau=\sin(\arg(\mu)-\omega\tau)$.
We can project the node perturbations $
\rho=(\rho_1,\dots,\rho_{N_0})^\top$,
$\theta=(\theta_1,\dots,\theta_{N_0})^\top$
onto the left singular vectors $\Psi^{(\alpha)}_{0}$ of $\mathbf{B}_1$, and the edge
perturbations $\eta=(\eta_1,\dots,\eta_{N_1})^\top$, $\varphi=(\varphi_1,\dots,\varphi_{N_1})^\top$
onto the right singular vectors $\Psi^{(\alpha)}_{1}$. Let $\Lambda_\alpha$ denote the singular value associated with the pair
$(\Psi_0^{(\alpha)},\Psi_1^{(\alpha)})$. Then for each $\alpha$ the mode evolves
independently according to
\begin{equation}
\left\{
\begin{aligned}
\frac{d}{dt}\hat\rho_\alpha
&=
-2\sigma_{\Re}\hat\rho_\alpha
-\Lambda_\alpha |\mu|r_0^{m-1}
\Bigl[
ma_\tau\,\hat\eta_\alpha^{\tau}
-b_\tau\,\hat\varphi_\alpha^{\tau}
\Bigr],
\\[4pt]
\frac{d}{dt}\hat\theta_\alpha
&=
-2\beta_{\Im}\frac{\sigma_{\Re}}{\beta_{\Re}}\hat\rho_\alpha
-\Lambda_\alpha |\mu|r_0^{m-1}
\Bigl[
mb_\tau\,\hat\eta_\alpha^{\tau}
+a_\tau\,\hat\varphi_\alpha^{\tau}
\Bigr],
\\[4pt]
\frac{d}{dt}\hat\eta_\alpha
&=
-2\sigma_{\Re}\hat\eta_\alpha
-\Lambda_\alpha |\mu|r_0^{m-1}
\Bigl[
ma_\tau\,\hat\rho_\alpha^\tau
-b_\tau\,\hat\theta_\alpha^\tau
\Bigr],
\\[4pt]
\frac{d}{dt}\hat\varphi_\alpha
&=
-2\beta_{\Im}\frac{\sigma_{\Re}}{\beta_{\Re}}\hat\eta_\alpha
-\Lambda_\alpha |\mu|r_0^{m-1}
\Bigl[
mb_\tau\,\hat\rho_\alpha^\tau
+a_\tau\,\hat\theta_\alpha^\tau
\Bigr].
\end{aligned}
\right.
\label{eq:MSE_D1graph}
\end{equation}

As noted in the previous section, time delay significantly complicates the analysis of the stability of the synchronous state. In order to obtain a closed analytical treatment in terms of the Lambert W-function, we assume  $b_\tau=0$ and $a_\tau=1$, hence

\begin{equation}
\arg(\mu)-\omega\tau_p=2p\pi,
\quad \tau_p>0 \text{ and } p\in\mathbb{Z}\,.
\label{eq:condiionHodeLapTau2}
\end{equation}

Note that, we are again limiting ourselves here to the case $a_\tau=1$. The case $a_\tau=-1$ follows naturally and always corresponds to unstable phase modes. From system Eq.~\eqref{eq:MSE_D1graph} we can extract an amplitude variational subsystem
\begin{equation}
\left\{
\begin{aligned}
\frac{d}{dt}{\hat{\rho}}_{\alpha}(t)
&=
-2\sigma_{\Re}\,\hat{\rho}_{\alpha}(t)
-m\Lambda_{\alpha} |\mu|r_0^{m-1}\,\hat{\eta}_{\alpha}^{\tau_p},\\[1mm]
\frac{d}{dt}{\hat{\eta}}_{\alpha}(t)
&=
-2\sigma_{\Re}\,\hat{\eta}_{\alpha}(t)
-m\Lambda_{\alpha} |\mu|r_0^{m-1}\,\hat{\rho}_{\alpha}^{\tau_p}.
\end{aligned}
\right.
\label{eq:amp-subsystem}
\end{equation}
By looking for exponential solutions
$\hat{\rho}_{\alpha}(t)=\hat{\rho}_{\alpha}(0)\,e^{\lambda t}$ and $\hat{\eta}_{\alpha}(t)=\hat{\eta}_{\alpha}(0)\,e^{\lambda t}$ we obtain the characteristic equation
\begin{equation}
\bigl(\lambda+2\sigma_{\Re}\bigr)^2
-
(m\Lambda_{\alpha} r_0^{m-1} |\mu|)^2e^{-2\lambda\tau_p}
=0\, ,
\label{eq:amp-chargraph}
\end{equation}
namely
\begin{widetext}
\begin{equation}
\left[\lambda+2\sigma_{\Re}-
(m\Lambda_{\alpha} r_0^{m-1} |\mu|)e^{-\lambda\tau_p}\right]\left[\lambda+2\sigma_{\Re}+
(m\Lambda_{\alpha} r_0^{m-1} |\mu|)e^{-\lambda\tau_p}\right]
=0\, .
\label{eq:amp-chargraph2}
\end{equation}
\end{widetext}
By leveraging again on the Lambert W-function and its property, one can conclude that the solution with the largest real part will be selected among the following ones
\begin{equation}
\begin{aligned}
\lambda_{\alpha,0}^{(\mathrm{amp})}
&=
-2\sigma_{\Re}
+
\frac{1}{\tau_p}
W_0\!\left(\pm \tau_p mr_0^{m-1}\Lambda_{\alpha}|\mu|\,e^{2\sigma_{\Re}\tau_p}
\right).
\label{eq:amp-lambert1}
\end{aligned}
\end{equation}
The remaining equations in Eq.~\eqref{eq:MSE_D1graph} involve the evolution of the phases $(\hat{\theta}_{\alpha},\hat{\phi}_{\alpha})$
\begin{equation}
\left\{
\begin{aligned}
\frac{d}{dt}{\hat{\theta}}_{\alpha}(t)
&=
-2\beta_{\Im}\frac{\sigma_{\Re}}{\beta_{\Re}}\hat{\rho}_\alpha-\Lambda_{\alpha}|\mu|r_0^{m-1}\,\hat{\phi}_{\alpha}^{\tau_p},\\[1mm]
\frac{d}{dt}{\hat{\phi}}_{\alpha}(t)
&=
-2\beta_{\Im}\frac{\sigma_{\Re}}{\beta_{\Re}}\hat{\eta}_\alpha-\Lambda_{\alpha}|\mu|r_0^{m-1}\,\hat{\theta}_{\alpha}^{\tau_p}.
\end{aligned}
\right.
\label{eq:phase-subsystem}
\end{equation}
Once $\hat{\rho}_\alpha$ and $\hat{\eta}_\alpha$ have been determined, the latter system is non-autonomous and thus to solve it we first need to solve the homogeneous part
\begin{equation}
\left\{
\begin{aligned}
\frac{d}{dt}{\hat{\theta}}_{\alpha}(t)
&=-\Lambda_{\alpha}|\mu|r_0^{m-1}\,\hat{\phi}_{\alpha}^{\tau_p},\\[1mm]
\frac{d}{dt}{\hat{\phi}}_{\alpha}(t)
&=
-\Lambda_{\alpha}|\mu|r_0^{m-1}\,\hat{\theta}_{\alpha}^{\tau_p}.
\end{aligned}
\right.
\label{eq:phase-subsystemhom}
\end{equation}
whose characteristic equation is
\begin{equation}
\lambda^2 - (\Lambda_{\alpha}r_0^{m-1} |\mu|e^{-\lambda\tau_p})^2=0\, .
\label{eq:phase-char}
\end{equation}
Its solution with the largest real part is
\begin{equation}
\lambda_{\alpha,0}^{(\mathrm{ph})}
=
\frac{1}{\tau_p}
W_0\!\left( \tau_p \Lambda_{\alpha}r_0^{m-1}|\mu|
\right)\, .
\label{eq:phase-lambertgraph}
\end{equation}

Let us observe that the argument of the Lambert W-function is positive and thus $\lambda_{\alpha,0}^{(\mathrm{ph})}>0$, for all $\alpha>1$. Consequently, the perturbations in the phases exhibit an exponential growth and thus \textrm{GTDS} cannot emerge.

Let us now consider a $2$-dimensional simplicial (or cell) complex in which each decoupled topological
signal evolves according to the Stuart-Landau dynamics. In this case, the
delayed Dirac-coupled model reads
\begin{widetext}
\begin{equation}
\left\{
\begin{aligned}
\frac{d}{dt}u_j(t)
&=
\sigma\,u_j(t)-\beta\,u_j(t)\lvert u_j(t)\rvert^2
-\mu\sum_{\ell=1}^{N_1} \mathbf{B}_1(j,\ell)\,h(v_\ell^\tau),
\\[4pt]
\frac{d}{dt}v_\ell(t)
&=
\sigma\,v_\ell(t)-\beta\,v_\ell(t)\lvert v_\ell(t)\rvert^2
-\mu\sum_{j=1}^{N_0} \mathbf{B}_1^\top(\ell,j)\,h(u_j^\tau)
-\mu\sum_{r=1}^{N_2} \mathbf{B}_2(\ell,r)\,h(z_r^\tau),
\\[4pt]
\frac{d}{dt}z_r(t)
&=
\sigma\,z_r(t)-\beta\,z_r(t)\lvert z_r(t)\rvert^2
-\mu\sum_{\ell=1}^{N_1} \mathbf{B}_2^\top(r,\ell)\,h(v_\ell^\tau),
\end{aligned}
\right.
\label{eq:SL_dirac_delayed}
\end{equation}
\end{widetext}
where $u_j(t)$, $v_\ell(t)$, and $z_r(t)$ are complex variables supported,
respectively, on nodes, edges, and triangles and describing the complex amplitude of the oscillators. 
We now study the stability of the synchronous state $ \mathbf{1_{\mathcal{N}}}\otimes s(t)$ by
linearizing Eq.~\eqref{eq:SL_dirac_delayed} about this periodic orbit.
We introduce again heterogeneous perturbations in amplitude and phase of topological signals by writing
$u_j(t)=\mathbf{s}(t)\bigl(1+\rho_j(t)\bigr)e^{i\theta_j(t)}$, $v_\ell(t)=\mathbf{s}(t)\bigl(1+\eta_\ell(t)\bigr)e^{i\varphi_\ell(t)}$ and $z_r(t)=\mathbf{s}(t)\bigl(1+\xi_r(t)\bigr)e^{i\psi_r(t)},
$ where $
\rho_j$, $\theta_j$, $\eta_\ell$, $\varphi_\ell$, $\xi_r$ and $\psi_r$
are small real-valued perturbations.

By substituting these expressions into the delayed equations and retaining only the
linear terms yields for the node dynamics
\begin{equation}
\left\{
\begin{aligned}
\frac{d}{dt}\rho_j
&=
-2\sigma_{\Re}\rho_j+\\&
-|\mu|r_0^{m-1}\sum_{\ell=1}^{N_1}\mathbf{B}_1(j,\ell)
\Bigl[
ma_\tau\,\eta^\tau_\ell-b_\tau\,\varphi^\tau_\ell
\Bigr],
\\[4pt]
\frac{d}{dt}\theta_j
&=
-2\beta_{\Im}\frac{\sigma_{\Re}}{\beta_{\Re}}\rho_j+\\&
-|\mu|r_0^{m-1}\sum_{\ell=1}^{N_1}\mathbf{B}_1(j,\ell)
\Bigl[
mb_\tau\,\eta^\tau_\ell+a_\tau\,\varphi^\tau_\ell
\Bigr],
\end{aligned}
\right.
\label{eq:linear_nodesD}
\end{equation}
Applying the same linearization procedure to the edge and triangle variational equations
gives
{\small
\begin{equation}
\left\{
\begin{aligned}
\frac{d}{dt}\eta_\ell
&=
-2\sigma_{\Re}\eta_\ell
\\
&\quad
-|\mu|r_0^{m-1}
\sum_{j=1}^{N_0} \mathbf{B}_1^\top(\ell,j)
\Bigl[
ma_\tau\,\rho^\tau_j-b_\tau\,\theta^\tau_j
\Bigr]
\\
&\quad
-|\mu|r_0^{m-1}
\sum_{r=1}^{N_2} \mathbf{B}_2(\ell,r)
\Bigl[
ma_\tau\,\xi^\tau_r-b_\tau\,\psi^\tau_r
\Bigr],
\\[3pt]
\frac{d}{dt}\varphi_\ell
&=
-2\beta_{\Im}\frac{\sigma_{\Re}}{\beta_{\Re}}\eta_\ell
\\
&\quad
-|\mu|r_0^{m-1}
\sum_{j=1}^{N_0} \mathbf{B}_1^\top(\ell,j)
\Bigl[
mb_\tau\,\rho^\tau_j+a_\tau\,\theta^\tau_j
\Bigr]
\\
&\quad
-|\mu|r_0^{m-1}
\sum_{r=1}^{N_2} \mathbf{B}_2(\ell,r)
\Bigl[
mb_\tau\,\xi^\tau_r+a_\tau\,\psi^\tau_r
\Bigr].
\end{aligned}
\right.
\label{eq:linear_edgesD}
\end{equation}
}
and
{\small
\begin{equation}
\left\{
\begin{aligned}
\frac{d}{dt}\xi_r
&=
-2\sigma_{\Re}\xi_r+
\\&
-|\mu|r_0^{m-1}\sum_{\ell=1}^{N_1} \mathbf{B}_2^\top(r,\ell)
\Bigl[
ma_\tau\,\eta^\tau_\ell-b_\tau\,\varphi^\tau_\ell
\Bigr],
\\[4pt]
\frac{d}{dt}\psi_r
&=
-2\beta_{\Im}\frac{\sigma_{\Re}}{\beta_{\Re}}\xi_r+\\&
-|\mu|r_0^{m-1}\sum_{\ell=1}^{N_1} \mathbf{B}_2^\top(r,\ell)
\Bigl[
mb_\tau\,\eta^\tau_\ell+a_\tau\,\varphi^\tau_\ell
\Bigr].
\end{aligned}
\right.
\label{eq:linear_trianglesD}
\end{equation}}

The systems Eq.~\eqref{eq:linear_nodesD},~\eqref{eq:linear_edgesD} and Eq.~\eqref{eq:linear_trianglesD} are coupled, however Dirac decomposition Eq.~\eqref{eq:spinordecomposition} reduces the problem to two independent systems, by considering the decomposition $(\eta_\ell,\varphi_\ell)^\top = (\eta_\ell^{harm},\varphi^{harm}_\ell)^\top + (\eta_\ell^{[1]},\varphi^{[1]}_\ell)^\top+(\eta_\ell^{[2]},\varphi^{[2]}_\ell)^\top$. Let us first focus on the case associated
with the node-edge coupling through $\mathbf{B}_1$ which yields 
\begin{equation}
\left\{
\begin{aligned}
\frac{d}{dt}\rho_j
&=
-2\sigma_{\Re}\rho_j+\\&
-|\mu|r_0^{m-1}\sum_{\ell=1}^{N_1} \mathbf{B}_1(j,\ell)
\Bigl[
ma_\tau\,\eta_\ell^{\tau[1]}-b_\tau\,\varphi_\ell^{\tau[1]}
\Bigr],
\\[4pt]
\frac{d}{dt}\theta_j
&=
-2\beta_{\Im}\frac{\sigma_{\Re}}{\beta_{\Re}}\rho_j+\\&
-|\mu|r_0^{m-1}\sum_{\ell=1}^{N_1} \mathbf{B}_1(j,\ell)
\Bigl[
mb_\tau\,\eta_\ell^{\tau[1]}+a_\tau\,\varphi_\ell^{\tau[1]}
\Bigr],
\\[4pt]
\frac{d}{dt}\eta_\ell^{[1]}
&=
-2\sigma_{\Re}\eta_\ell^{[1]}+\\&
-|\mu|r_0^{m-1}\sum_{j=1}^{N_0} \mathbf{B}_1^\top(\ell,j)
\Bigl[
ma_\tau\,\rho^\tau_j-b_\tau\,\theta^\tau_j
\Bigr],
\\[4pt]
\frac{d}{dt}\varphi_\ell^{[1]}
&=
-2\beta_{\Im}\frac{\sigma_{\Re}}{\beta_{\Re}}\eta_\ell^{[1]}+\\&
-|\mu|r_0^{m-1}\sum_{j=1}^{N_0} \mathbf{B}_1^\top(\ell,j)
\Bigl[
mb_\tau\,\rho^\tau_j+a_\tau\,\theta^\tau_j
\Bigr].
\end{aligned}
\right.
\label{eq:first_dirac_sector}
\end{equation}
We can project the node perturbations
$\rho=(\rho_1,\dots,\rho_{N_0})^\top$, $
\theta=(\theta_1,\dots,\theta_{N_0})^\top$
 onto the left singular vectors $\Psi^{(\alpha)}_{0}$ of $\mathbf{B}_1$, and the edge
perturbations
$\eta^{[1]}=(\eta_1^{[1]},\dots,\eta_{N_1}^{[1]})^\top$, $\varphi^{[1]}=(\varphi_1^{[1]},\dots,\varphi^{[1]}_{N_1})^\top$ onto the right singular vectors $\Psi^{(\alpha)}_{1}$. By defining
$\Lambda_\alpha$ as the singular value associated with the pair
$(\Psi_0^{(\alpha)},\Psi_1^{(\alpha)})$, each $\alpha-$mode evolves
independently according to
\begin{equation}
\label{eq:MSE_nodelink}
\left\{
\begin{aligned}
\frac{d}{dt}\hat{\rho}_\alpha
&=
-2\sigma_{\Re}\hat{\rho}_\alpha
-\Lambda_\alpha |\mu|\,r_0^{m-1}
\Bigl(
m a_\tau\,\hat{\eta}_\alpha^{[1]\tau}
-b_\tau\,\hat{\varphi}_\alpha^{[1]\tau}
\Bigr),
\\[4pt]
\frac{d}{dt}\hat{\theta}_\alpha
&=
-2\beta_{\Im}
\frac{\sigma_{\Re}}{\beta_{\Re}}
\hat{\rho}_\alpha
-\Lambda_\alpha |\mu|\,r_0^{m-1}
\Bigl(
m b_\tau\,\hat{\eta}_\alpha^{[1]\tau}
+a_\tau\,\hat{\varphi}_\alpha^{[1]\tau}
\Bigr),
\\[4pt]
\frac{d}{dt}\hat{\eta}_\alpha^{[1]}
&=
-2\sigma_{\Re}\hat{\eta}_\alpha^{[1]}
-\Lambda_\alpha |\mu|\,r_0^{m-1}
\Bigl(
m a_\tau\,\hat{\rho}_\alpha^\tau
-b_\tau\,\hat{\theta}_\alpha^\tau
\Bigr),
\\[4pt]
\frac{d}{dt}\hat{\varphi}_\alpha^{[1]}
&=
-2\beta_{\Im}
\frac{\sigma_{\Re}}{\beta_{\Re}}
\hat{\eta}_\alpha^{[1]}
-\Lambda_\alpha |\mu|\,r_0^{m-1}
\Bigl(
m b_\tau\,\hat{\rho}_\alpha^\tau
+a_\tau\,\hat{\theta}_\alpha^\tau
\Bigr).
\end{aligned}
\right.
\end{equation}
If we introduce the vector $
Y_{\alpha}=
\begin{pmatrix}
\hat\rho_\alpha,
\hat\theta_\alpha,
\hat\eta_\alpha^{[1]},
\hat\varphi_\alpha^{[1]}
\end{pmatrix}^\top$ and the matrices
\[
\begin{aligned}
J_f&=
\begin{pmatrix}
-2\sigma_{\Re} & 0\\
-2\beta_{\Im}\dfrac{\sigma_{\Re}}{\beta_{\Re}} & 0
\end{pmatrix},
\quad
J_0^\tau=
|\mu|r_0^{m-1}\begin{pmatrix}
ma_\tau & -b_\tau\\
mb_\tau & a_\tau
\end{pmatrix},
\end{aligned}
\]
then Eq.~\eqref{eq:MSE_nodelink} can be rewritten in compact form as
\begin{equation}
\frac{d}{dt}Y_{\alpha}
=
\begin{pmatrix}
J_f & 0\\
0 & J_f
\end{pmatrix}
Y_{\alpha}
-
\Lambda_\alpha
\begin{pmatrix}
0 & J_0^\tau\\
J_0^\tau & 0
\end{pmatrix}
Y_{\alpha}(t-\tau).
\label{eq:compact_modal}
\end{equation}
Eq.~\eqref{eq:compact_modal} is the delayed variational equation governing
the stability of the synchronized Stuart-Landau state associated with the node-edge coupling. An analogous derivation can be carried out
with the edge-triangle coupling through $\mathbf{B}_2$.

If we consider again $b_\tau=0$ and $a_\tau=1$, then the analysis simplifies and we can conclude in the same way as for the system Eq.~\eqref{eq:MSE_D1graph}. Consequently, the \textrm{GTDS} cannot emerge for every time delay $\tau_p$.

We note that the same conclusion is obtained if delayed interactions are imposed only from simplexes, or cells, of dimension $k$ toward simplexes, or cells, of dimensions $k-1$ and $k+1$. To illustrate this point more clearly, let us consider a variant of the model Eq.~\eqref{eq:SL_dirac_delayedgraph} in which the second equation associated to links, is written without delayed interactions, namely
\begin{equation}
\left\{
\begin{aligned}
\frac{d}{dt}u_j(t)
&=
\sigma\,u_j(t)-\beta\,u_j(t)\lvert u_j(t)\rvert^2+\\&
-\mu\sum_{\ell=1}^{N_1} \mathbf{B}_1(j,\ell)\,h(v_\ell(t-\tau)),
\\[4pt]
\frac{d}{dt}v_\ell(t)
&=
\sigma\,v_\ell(t)-\beta\,v_\ell(t)\lvert v_\ell(t)\rvert^2+\\&
-\mu\sum_{j=1}^{N_0} \mathbf{B}_1^\top(\ell,j)\,h(u_j(t)),
\end{aligned}
\right.
\label{eq:SL_dirac_delayedgraph_v2}
\end{equation}

After projection onto the left and right singular vectors of the incidence
matrix $\mathbf{B}_1$, the asymmetric delay in Eq.~\eqref{eq:SL_dirac_delayedgraph_v2} yields the following
modal system:
\begin{equation}
\left\{
\begin{aligned}
\frac{d}{dt}\hat{\rho}_{\alpha}
={}&
-2\sigma_{\Re}\hat{\rho}_{\alpha}
-
\Lambda_\alpha|\mu|r_0^{m-1}
\left[
m a_{\tau}\hat{\eta}_{\alpha}^\tau
-
b_{\tau}\hat{\varphi}_{\alpha}^\tau
\right],
\\[2mm]
\frac{d}{dt}\hat{\theta}_{\alpha}
={}&
-2\beta_{\mathrm{Im}}
\frac{\sigma_{\Re}}{\beta_{\Re}}
\hat{\rho}_{\alpha}
-
\Lambda_\alpha|\mu|r_0^{m-1}
\left[
m b_{\tau}\hat{\eta}_{\alpha}^\tau
+
a_{\tau}\hat{\varphi}_{\alpha}^\tau
\right],
\\[2mm]
\frac{d}{dt}\hat{\eta}_{\alpha}
={}&
-2\sigma_{\Re}\hat{\eta}_{\alpha}
-
\Lambda_\alpha|\mu|r_0^{m-1}
\left[
m a_{0}\hat{\rho}_{\alpha}
-
b_{0}\hat{\theta}_{\alpha}
\right],
\\[2mm]
\frac{d}{dt}\hat{\varphi}_{\alpha}
={}&
-2\beta_{\mathrm{Im}}
\frac{\sigma_{\Re}}{\beta_{\Re}}
\hat{\eta}_{\alpha}
-
\Lambda_\alpha|\mu|r_0^{m-1}
\left[
m b_{0}\hat{\rho}_{\alpha}
+
a_{0}\hat{\theta}_{\alpha}
\right],
\end{aligned}
\right.
\label{eq:asymmetric_modal_system}
\end{equation}
where the coefficients associated with the instantaneous interaction are
$
a_{0}=\cos\!\left(\arg(\mu)\right)$ and $b_{0}=\sin\!\left(\arg(\mu)\right)$.

Nevertheless, even in this asymmetric-delay setting, the four variational equations remain coupled, and a general analytical treatment is still not available without additional assumptions. In order to obtain a tractable expression in terms of the Lambert $W$-function, one has to impose conditions analogous to those used above. In particular, since the last two equations do not contain delayed coupling terms, the coefficient $\mu$ must be chosen real in order to decouple the amplitude and phase variational dynamics. Under these assumptions, the phase variational subsystem leads to characteristic roots expressed through the principal branch of the Lambert W-function with a positive argument. Hence, the phase perturbations cannot be damped, and the topological oscillator phases cannot synchronize. This prevents the emergence of \textrm{GTDS}.

An analogous argument applies to simplicial complexes or cell complexes of dimension $K=2$, when the delay is present only in one component of the coupling function $\mathbf{H}_\tau$. Therefore, introducing delay in only one part of the Dirac coupling does not restore \textrm{GTDS}; the instability of the phase modes remains.

\subsection{Numerical results}
\label{sec:4B}

\begin{figure}
    \centering
\includegraphics[width=1\linewidth]{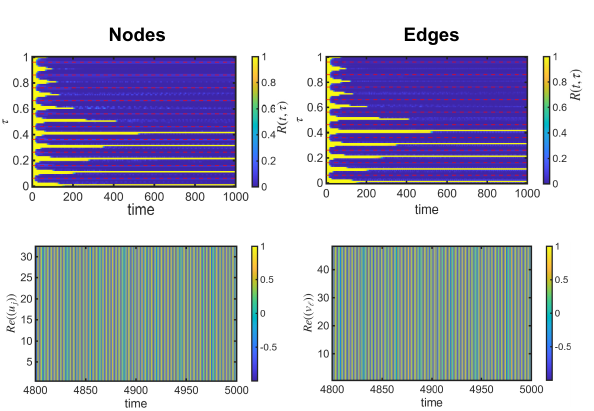}
    \caption{\textbf{GTDS in $1$-simplicial complex induced by time-delayed Dirac interactions}. The left and right columns refer, respectively, to topological oscillators supported on nodes and edges. The top row shows the generalized order parameters as functions of time for homogeneous delays $\tau\in[0,1]$. The horizontal red dashed lines mark the discrete admissible delays $\tau_p$ given by Eq.~\eqref{eq:condiionHodeLapTau2}. The bottom row shows the time evolution of the real part of the complex amplitude of topological oscillators on the nodes and edges with delay $\tau\simeq0.21\ne\tau_p$.The remaining parameters are $\sigma=1+10\pi i$, $\beta=1$, $\mu=0.2-0.5i$, and $a=1$.}
\label{fig:figurestimedelayedDirac}
\end{figure}

\begin{figure*}
    \centering
        \includegraphics[width=0.8\linewidth]{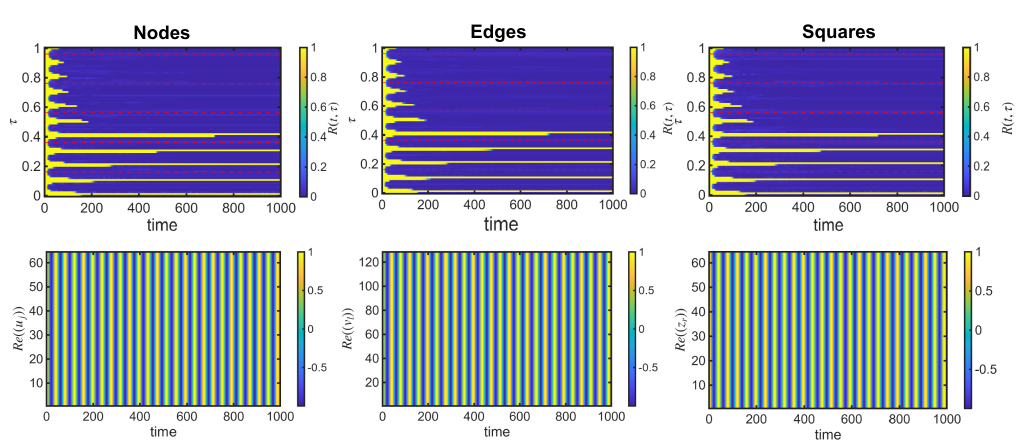}

\caption{\textbf{GTDS in a torus-shaped cell complex induced by time-delayed Dirac interactions.}
The left, middle, and right columns correspond, respectively, to the Stuart-Landau oscillators supported on the nodes, edges, and square cells of the complex. The top row shows the corresponding generalized order parameters, as functions of time and of the homogeneous delay $\tau\in[0,1]$. The horizontal red dashed lines indicate the discrete admissible delay values $\tau_p$ determined by the delay condition. The bottom row displays the time evolution of the real parts of the complex amplitudes, for the oscillators supported on nodes, edges, and square cells, respectively. These panels were obtained for the non-admissible delay $\tau\simeq0.41$, with $\tau\neq\tau_p$, and illustrate the simultaneous coherent dynamics of the three topological signal dimensions. The toroidal cell complex consists of $N=64$ nodes, $E=128$ edges, and $S=64$ square cells. The remaining model parameters are $\sigma=1+10\pi i$, $\beta=1$, $\mu=0.2-0.5i$, and $a=1$.}
 
    \label{fig:figurestimedelayedtopologicalsignals5}
\end{figure*}

We now illustrate the effect of delayed Dirac interactions on the emergence of \textrm{GTDS} to support and complement the theory above introduced. The aim of the following simulations is therefore to show how this analytical obstruction appears in finite systems.

We first consider the $1$-dimensional simplicial complex (see left panel of Fig.~\ref{fig:illustrationUnderlyingnetwork}) composed of $N=32$ nodes and $E=48$ edges, each one supporting a Stuart-Landau oscillator coupled with the Dirac matrix and where homogeneous delay acts on the coupling term. We thus numerically solve the DDE --see Eq.~\eqref{eq:SL_dirac_delayedgraph} and we report the results in Fig.~\ref{fig:figurestimedelayedDirac}. 

Top panels display the order parameters calculated based on Eq.~\eqref{eq:OP} and associated with the node and edge topological oscillators as functions of time and of the delay $\tau \in [0,1]$. Dark blue points correspond to low values of the order parameter and thus to absence of synchronization, while yellow points are associated to values close to $1$ and thus to global synchronization. The dashed red curves indicate the admissible values of the delay obtained from condition Eq.~\eqref{eq:condiionHodeLapTau2} imposed to develop the analytical theory as we did in the case of Hodge-Laplace coupling, however in the present case those values of $\tau_p$ do not generate synchronization as shown above. The results presented in the top panels show that coherent responses, i.e., global synchronization, can occur for certain isolated values of the delay that do not correspond to $\tau_p$. Bottom panels of Fig.~\ref{fig:figurestimedelayedDirac} show the time evolution of the real part of the complex amplitudes of the  nodes and edges for $\tau \simeq 0.21$. These plots illustrate the \textrm{GTDS} for a value of $\tau \neq \tau_p$.  

\begin{figure}
    \centering
    \includegraphics[width=0.8\linewidth]{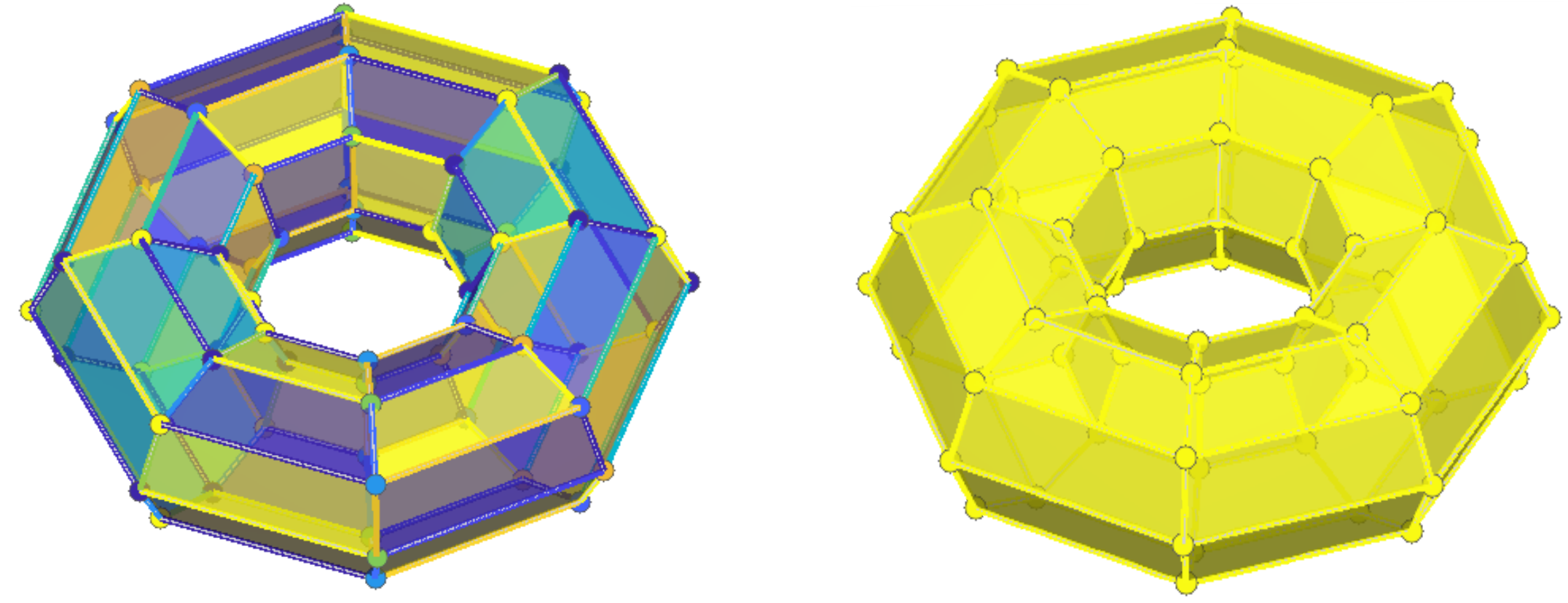}
    \caption{\textbf{Illustration of GTDS bserved for a non-admissible delay value.}
The panel on the left shows the dynamics of the complex cell at the final time $t_f=1000$ without delayed interactions, whilst the panel on the right corresponds to the state of the complex cell for $\tau \simeq0.41$.  The simulations were obtained using $\sigma = 1+10\pi i$, $\beta = 1$, $\mu = 0.2-0.5i$, $a=1$, $N =64$ and $E=128$ and $S=64$. The integration of the dynamics was achieved using $\tau\simeq0.41$. } 
    \label{fig:figurestimedelayedtopologicalsignals22}
\end{figure}

We then consider the $2$-cell complex shown in the right panel of Fig.~\ref{fig:illustrationUnderlyingnetwork}, i.e., a torus tessellated with squares, composed of $N=64$ nodes, $E=128$ edges, and $S=64$ $2$-dimensional cells (squares). The results of the numerical solution of the DDE--see Eq.~\eqref{eq:SL_dirac_delayedgraph} are shown in Fig.~\ref{fig:figurestimedelayedtopologicalsignals5}. Top panels display the order parameters associated with the nodes, edges, and squares. The same qualitative behavior as in the previous example of the network is observed: the admissible values of $\tau_p$ do not produce a synchronization for the delay values in the interval $[0,1]$. Bottom panels of Fig.~\ref{fig:figurestimedelayedtopologicalsignals5} show the time evolution of the real parts of the complex amplitudes for $\tau \simeq 0.41$. For this value of the delay, the dynamics synchronize across nodes, edges, and squares. This is further illustrated in Fig.~\ref{fig:figurestimedelayedtopologicalsignals22}, where we report the values of the real part of the complex amplitude of the oscillators of the cell complex with and without delayed interactions at time $t_f$. In the absence of delay, the final configuration displays heterogeneous values of the real part of the complex amplitude, whereas the delayed dynamics produce a \textrm{GTDS}.

\begin{figure*}
    \centering
    \includegraphics[width=0.8\linewidth]{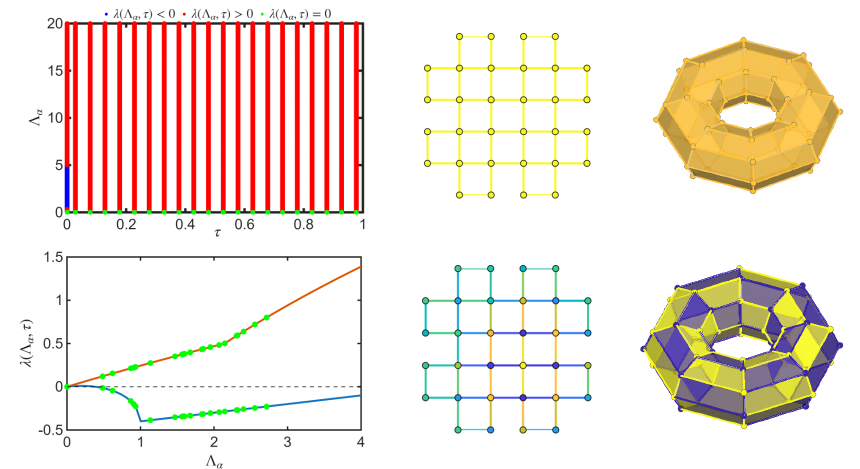}
    \caption{\textbf{Desynchronization induces by delayed Dirac interactions with $\tau=\tau_p$.} The panels on the left show the dispersion relation as a function of the singular values. The panel above calculates this dispersion relation for the possible discrete values of the delay and the possible values of the singular values. Red corresponds to a positive dispersion relation, and blue to a negative dispersion relation. The bottom panel shows the evolution of the dispersion relation as a function of the singular values of $\mathbf{B}_1$ in the absence of delay (blue curve and green points above the curve corresponding to the singular values of the matrix $\mathbf{B}_1$ of the  network) and for a delay $\tau = (\arg{\mu}+2\pi)/\omega$ (red curve with green points above for the singular values of the same network). The middle panels show the state of the  network at the final time; the colors represent the real parts of the complex amplitudes of the nodes and edges at the final time. The top panel is obtained with $\tau=0$, and the bottom one with $\tau=(\arg{\mu}+2\pi)/\omega$. The right-hand panels show the structures of a torus-shaped complex cell, where all $2$-cells are squares. The top panel corresponds to the case $\tau = 0$, and the bottom one to the case $\tau=(\arg{\mu}+2\pi)/\omega$. The other parameters used in the simulation are $\mu = 0.1–0.5i; \sigma = 0.5+20\pi i$, $\beta = 1$, and $a=2$. The  network is the same as the one used above, and the $2-$cell complex consists of 64 nodes, 128 edges, and 64 squares.
} 
    \label{fig:diractorusgraph}
\end{figure*}

A similar conclusion holds true when nonlinear interactions through the Dirac operator are considered: \textrm{GTDS} cannot emerge for the admissible delay values $\tau=\tau_p$ as illustrated in  Fig.~\ref{fig:diractorusgraph}. The upper-left panel shows the sign of the dispersion relation computed for the admissible discrete values of the delay and for the singular values of the incidence matrix. Red regions correspond to positive dispersion relation, while blue regions correspond to negative dispersion relation. Except for the zero singular value associated with the synchronous manifold (green dots), the delayed Dirac coupling produces positive dispersion relation. This agrees with the analytical prediction that the phase branch of the dispersion relation remains positive.
The lower-left panel of Fig.~\ref{fig:diractorusgraph} compares the dispersion relation without delay and with delay. In the absence of delay, the singular values of $\mathbf{B}_1$ may lie in a stable region of the dispersion relation, allowing the synchronized state to persist. When the delay is introduced, the dispersion relation becomes positive at the same singular values, and the synchronized state is destabilized. The middle and right panels confirm this prediction. For $\tau=0$, both the network and the torus-shaped cell complex display homogeneous final configurations. By contrast, for the delayed case, the final states become heterogeneous, showing that the delayed Dirac interaction destroys the synchronized state for $\tau=\tau_p$. Note that although the singular values of $\mathbf{B}_1$ and $\mathbf{B}_2$ associated with the cell complexes are not shown in the lower-left panel for the sake of clarity, they are indeed located within the stability region in the absence of delay.

\section{Conclusion}
\label{sec:conclusion}
So far, time-delayed interactions have been extensively studied in networked systems, where links model pairwise interactions and nodes describe the dynamics of individual units. In this work, we have explored the effect of such delays on the emergence of synchronization of topological oscillators supported on simplexes or cells and coupled through either the Hodge-Laplacian or the Dirac operator. We have shown that, for suitable choices of the model parameters, time delay can induce both \textrm{GTS} and \textrm{GTDS}, whereas for the same parameter values synchronization does not occur in the absence of delay. Moreover, we have identified a class of delay values for which a complete analytical stability analysis becomes possible, and we have shown in particular that within the analytically tractable class, delay can either promote or suppress GTS depending on the model parameters and on the Hodge-Laplacian spectrum, whereas the same class prevents GTDS.

Nevertheless, because of the complexity introduced by delayed interactions, several assumptions had to be imposed, most notably the restriction to homogeneous delays. This work therefore opens several promising directions for future research. First, it would be natural to relax the assumption of homogeneous delay and investigate the case of heterogeneous delays, which are more realistic in many applications. Another important direction would be to study the impact of delayed higher-order interactions on other collective phenomena, such as cluster synchronization or pattern formation.

\begin{acknowledgments}
This work is supported by the European Union - NextGenerationEU - National Recovery and Resilience Plan (Piano Nazionale di Ripresa e Resilienza, PNRR), project `SoBigData.it - Strengthening the Italian RI for Social Mining and Big Data Analytics' - Grant IR0000013 (n. 3264, 28/12/2021) (\url{https://pnrr.sobigdata.it/}).
\end{acknowledgments}

\section*{Competing interests}
The authors declare no competing interests.

\nocite{*}
\bibliography{mybib}

\end{document}